\documentclass[aps,prx,twocolumn,superscriptaddress,amsmath,amssymb,longbibliography,floatfix,nofootinbib]{revtex4-2}
\usepackage{graphicx}
\usepackage{xcolor}
\usepackage{amsthm}

\usepackage[colorlinks=true,linkcolor=blue,citecolor=blue,urlcolor=blue]{hyperref}

\newcommand{\ad}{\hat a^\dagger}
\newcommand{\aop}{\hat a}
\newcommand{\bop}{\hat b}
\newcommand{\bd}{\hat b^\dagger}
\newtheorem{theorem}{Theorem}
\begin{document}
\title{Dynamical Casimir emission from Floquet-driven Landau polaritons:\\ Gauge invariance forbids a superradiant phase transition}
\author{S. Hughes}
\affiliation{Department of Physics, Engineering Physics and Astronomy, Queen's University, Kingston, Ontario K7L 3N6, Canada}
\date{\today}

\begin{abstract}
Ultrastrong light--matter coupling of linear systems, such as Landau polaritons formed by a two-dimensional electron gas in a terahertz cavity, is described by the bosonic Hopfield model, and gauge invariance forbids a superradiant phase transition (SRPT). Modulating the magnetic field alone has been predicted to enhance the cycle-averaged coupling and induce a Floquet SRPT. We develop an exact theory of periodically driven Landau polaritons. Including the electric field induced by the modulated flux (Faraday's law), the problem reduces exactly, by Kohn's theorem, to the electronic center-of-mass mode coupled to the cavity, whose Hamiltonian is a sum of squares at every instant. A theorem of Krein then guarantees stability to all orders above a computable drive frequency, the Krein frequency, for which we give a self-contained proof. Below it, the system is stable except within resonance tongues where the drive creates pairs from the vacuum, the dynamical Casimir effect. At high drive frequency the electrons experience the mean field plus a ponderomotive trap (the Fock--Darwin problem), not an enhanced coupling. A gauge theorem excludes a Floquet photon condensate for any band structure, including Kane nonparabolicity, for which we derive a generalized Thomas--Reiche--Kuhn (TRK) sum rule. The predicted transition arises from averaging the coupling in a fixed Landau-level basis, which yields an effective model that violates the TRK sum rule; under driving, gauge invariance also requires the induced Faraday field. The rule to modulate and average the minimal-coupling Hamiltonian, not the parameters of its projection, applies to any linear, minimally coupled system. Instead of an SRPT, magnetic modulation produces dynamical Casimir emission: Faraday-driven polariton--magnetoplasmon pairs at low threshold, and two-color photon pairs with up to 5--10~dB of two-mode squeezing.
\end{abstract}
\maketitle

\section{Introduction}
\label{sec:intro}

Ultrastrong coupling (USC), in which the light--matter coupling becomes a sizable fraction of the bare frequencies, has become an experimental reality across platforms, from superconducting circuits to semiconductor microcavities and plasmonic and terahertz (THz) metamaterials~\cite{Ciuti2005,FriskKockum2019,FornDiaz2019,Niemczyk2010,Yoshihara2017,Gunter2009,Bayer2017}. A prominent example is the two-dimensional electron gas (2DEG) in a THz cavity. In a perpendicular magnetic field, the cyclotron transition of the electrons couples to the cavity mode to form Landau polaritons, with normalized couplings well beyond unity in the most extreme samples~\cite{Hagenmuller2010,Scalari2012,Zhang2016,Li2018,Bayer2017}. Landau polaritons have become a test bed for USC and for the broader program of controlling matter with cavity vacuum fields~\cite{GarciaVidal2021,Schlawin2022,Bloch2022}, including magnetotransport~\cite{ParaviciniBagliani2019} and the breakdown of topological protection by cavity vacuum fields~\cite{Appugliese2022}.

The collective response of Landau polaritons is described by the bosonic Hopfield model~\cite{Hopfield1958,Ciuti2005,Hagenmuller2010}. It is quadratic in the field operators and therefore linear, and it has an exact classical counterpart. When the light--matter coupling is constructed gauge invariantly~\cite{Garziano2020,DeBernardis2018,DiStefano2019,Gustin2023}, its spectra follow from classical Maxwell optics~\cite{ChangPRA2025,Muniain2025}. This remains true for real cavities with finite losses (open cavities), provided the gauge corrections and the cavity--bath coupling are treated consistently~\cite{Hughes2024}, including dissipation and input--output in the USC regime~\cite{GustinPRL2025,LSA2026}. The quantum Rabi model of a single two-level system is different. It is nonlinear even in the vacuum, and its Floquet driving produces genuinely quantum effects, including photon generation~\cite{AkbariPRL2025,AkbariFloquet2026}.

A second hallmark of collective USC is the superradiant phase transition (SRPT). In the Dicke model, $N$ dipoles coupled to a cavity mode acquire a macroscopic photon field in the ground state once the collective coupling exceeds a critical value~\cite{Dicke1954,HeppLieb1973,WangHioe1973,EmaryBrandes2003}. The cavity then holds a static, coherent field even in the absence of any drive, $\langle\aop\rangle\neq0$, and a symmetry of the normal (vacuum-like) state is spontaneously broken. In theoretical models, the transition is usually signaled by \emph{a polariton frequency that softens to zero at the critical coupling}. Beyond it, the normal state becomes unstable, and the energy as a function of the cavity field develops minima at a nonzero field. A nonequilibrium analog has been realized with atoms in a driven optical cavity~\cite{Baumann2010}. For electric-dipole coupling to a uniform cavity field, however, the equilibrium SRPT is forbidden~\cite{Rzazewski1975,Nataf2010,Andolina2019,DeBernardis2018,DiStefano2019,Garziano2020}. The original prediction relied on neglecting the diamagnetic ($A^2$) term, or equivalently the $P^2$ (dipole self-energy) term in the multipolar gauge. Gauge invariance fixes the strength of that term relative to the paramagnetic coupling; for a parabolic band this relation is the Thomas--Reiche--Kuhn (TRK) sum rule. When the term is retained with this strength, the paramagnetic and diamagnetic static responses cancel, and the transition is forbidden. Equivalently, a passive, causal medium cannot have a negative static response~\cite{Bamba2014}, a fundamental requirement from electromagnetism.

The no-go theorem is not an artifact of how the Hamiltonian is written. In the Maxwell picture, a collection of linear oscillators in a cavity has no SRPT. For collective electric dipoles this remains true once dipole--dipole (lattice, local-field) effects are included; those effects must be added to the simple Hopfield model or built into an effective one, and a correctly used Hopfield description then agrees with full classical electromagnetic simulations~\cite{Bamba2014,Lamberto2025}. A transition requires a nonlinear coupling beyond the linear (lowest-order Holstein--Primakoff~\cite{HolsteinPrimakoff1940}) description of the matter, and whether such corrections survive the thermodynamic limit while preserving gauge invariance remains open~\cite{Lamberto2025}. Related instabilities in spatially varying cavity fields likewise originate in the matter, for example a magnetostatic (Condon) instability of a strongly correlated electron system~\cite{Andolina2020}. Proposals for superradiant quantum materials accordingly rely on interactions, spatial structure, or nonlinear electronic degrees of freedom~\cite{MazzaGeorges2019,Guerci2020,Andolina2020,Andolina2022}. Consistent with this picture, a recent experiment on graphene (Dirac) Landau polaritons, whose nonparabolic bands had been predicted to allow an SRPT, found no transition~\cite{Jochl2026}.

Can periodic driving provide a route around the no-go theorem? Floquet engineering has become a powerful tool for designing effective Hamiltonians in quantum materials~\cite{OkaKitamura2019,Bukov2015,Eckardt2015,Hubener2021}, and the driven Dicke model shows nonequilibrium transitions~\cite{Bastidas2012}. Floquet routes to superradiance have also been proposed for cold atoms in multimode cavities~\cite{Johansen2022}, for the Dicke model with a modulated qubit frequency~\cite{HuangTian2022}, and for cavity magnonics~\cite{Lin2026}. For a linear system, however, the expectation is clear. Modulating its parameters can produce parametric amplification, i.e., the drive feeds energy into pairs of oscillations whose frequencies add up to a multiple of the drive frequency, as 
in an optical parametric amplifier. It cannot produce an equilibrium-like condensate, because the TRK protection holds at every instant. 

A Floquet SRPT of Landau polaritons would therefore be a significant finding, all the more so if it could be induced by modulating the magnetic field alone. Wu \emph{et al.}~\cite{Wu2026} recently proposed exactly this. In their scheme, an off-resonant modulation $B(t)=B_0[1+\varepsilon\cos\omega_pt]$, i.e., one whose frequency $\omega_p$ lies far above the cyclotron and cavity frequencies ($\omega_p\gg\omega_0,\omega_{\rm cav}$), produces a dc enhancement of the cycle-averaged coupling, which Ref.~\cite{Wu2026} likens to optical rectification, and hence a ``quasiequilibrium'' SRPT above a critical depth $\varepsilon_c$. Their analysis uses the Hopfield model with time-dependent parameters. The coupling rate then depends nonlinearly on the field ($\Omega\propto\sqrt{|B|}$) and is averaged within a fixed Landau-level basis. The enhancement arises because the cycle average of the nonlinear function $\sqrt{|B(t)|}$ exceeds $\sqrt{B_0}$. As we show, this nonlinearity lies in the parameter dependence of a basis-dependent matrix element, not in the response of the system. The underlying dynamics is linear in the fields and contains no such enhancement.

Two physical ingredients are missing from a model that only modulates the parameters of the Hopfield Hamiltonian. First, by Faraday's law of induction ($\nabla\times\mathbf E=-\partial_t\mathbf B$), a time-varying magnetic flux always induces a circulating electric field, as in a transformer or a betatron. This field grows with distance from the axis of the flux and accelerates the electrons. Second, for a parabolic band, Kohn's theorem states that a spatially uniform field couples only to the center-of-mass (c.m.) motion of the electrons, which is unaffected by electron--electron interactions~\cite{Kohn1961,Brey1989,MaksymChakraborty1990,Peeters1990,Dobson1994,Rokaj2019,Rokaj2023}. 

Together, these two effects reduce the modulated problem exactly, without a Landau-level truncation, to the c.m.\ (Kohn) mode in $B(t)$ and its Faraday field, coupled to the cavity. 
This is a linear model with an exact classical counterpart, and its stability and emission can be determined exactly.
Both ingredients express gauge invariance in its most physical form. The electrons respond to the actual electric and magnetic fields, including the induced one, and the same dynamics follows from the classical Maxwell--Lorentz equations. Gauge invariance is therefore not a formal requirement here, but a statement about which forces act on the electrons.

\begin{figure*}[t]
\centering\includegraphics[width=\textwidth]{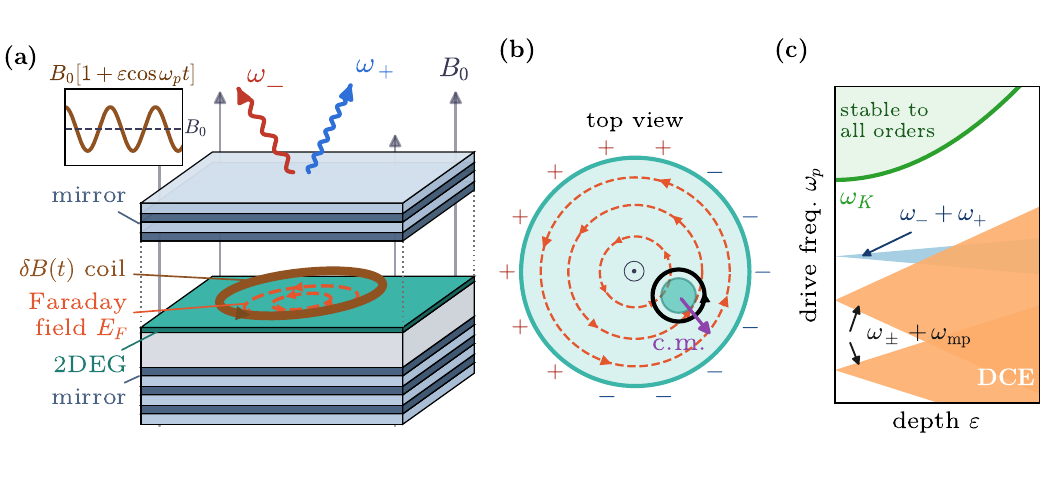}
\caption{Setup and summary. (a)~A 2DEG in a THz cavity (here, between two Bragg mirrors) in a static magnetic field $B_0$. A coil modulates the field, $B(t)=B_0[1+\varepsilon\cos\omega_pt]$ (inset). The modulated flux induces a circulating Faraday field $\mathbf E_F$ in the 2DEG plane, and pairs of photons at $\omega_-$ and $\omega_+$ leave the cavity. (b)~Top view: the Faraday field circulates about the flux axis ($\odot$) and drives the electronic center of mass (c.m.), whose cyclotron orbit and guiding-center drift are shown. Edge charges of a finite sample provide a restoring force (frequency $\omega_d$). (c)~Summary of the results. Above the Krein frequency $\omega_K(\varepsilon)$ the system is stable to all orders. Below it, the only instabilities are pair-creation (dynamical Casimir) tongues: polariton pairs at $\omega_-+\omega_+$ and Faraday-driven polariton--magnetoplasmon pairs at $\omega_\pm+\omega_{\rm mp}$. No superradiant phase transition occurs anywhere.}
\label{fig:schematic}
\end{figure*}

Figure~\ref{fig:schematic} sketches the setup and summarizes our results. Our main findings are as follows.
\begin{enumerate}
\item \emph{Positivity.} The Hamiltonian of the exact model is a sum of squares (the kinetic energy of the c.m.\ plus the cavity energy). Each term is the square of a Hermitian operator and can never be negative, so the energy is bounded below at every instant and no configuration of the fields can lower it without limit. This is the dynamical form of the TRK sum rule (Sec.~\ref{sec:positivity}).

\item \emph{Stability to all orders.} A theorem of Krein on periodically driven linear systems~\cite{Krein1955,GelfandLidskii1955,YakubovichStarzhinskii1975}, the multimode generalization of the familiar stability chart of the Mathieu equation,\footnote{The Mathieu equation, $\ddot x+\omega^2[1+h\cos\omega_pt]\,x=0$, describes an oscillator whose stiffness is modulated periodically. Its stability chart shows the regions (instability tongues) of the $(h,\omega_p)$ plane in which solutions grow exponentially; the widest starts at $\omega_p=2\omega$.} guarantees stability to all orders whenever the drive frequency $\omega_p$ exceeds a threshold that we call the Krein frequency, $\omega_K(\varepsilon)$. It is obtained from the instantaneous Hamiltonian without solving the dynamics. We give a self-contained proof (Theorem~\ref{thm:krein}).

\item \emph{Pair creation is the only instability.} For $\omega_p<\omega_K$, stability can be lost only at sum resonances, which create pairs of real excitations from the vacuum. This is the dynamical Casimir effect (DCE)~\cite{Moore1970,FullingDavies1976,Dodonov2010,Nation2012,Wilson2011,Lahteenmaki2013,DeLiberato2007,Hagenmuller2016}. In a linear system the emission is seeded by vacuum fluctuations alone; a classical counterpart left at rest stays at rest.

\item \emph{High-frequency limit.} For $\omega_p\gg\omega_\pm,\omega_0$ the electrons experience the mean magnetic field plus a ponderomotive trap from the Faraday field. This is the Fock--Darwin problem of quantum dots~\cite{Fock1928,Darwin1931},\footnote{The Fock--Darwin problem describes the energy levels of a charged particle in a two-dimensional parabolic potential subjected to a perpendicular magnetic field. Its spectrum interpolates between the harmonic-oscillator levels at zero field and the Landau levels at high field.} which is stable and obeys the TRK sum rule. The light--matter coupling in this limit is the same as for a static field, fixed by the electron charge and mass, and is not enhanced by the drive (Sec.~\ref{sec:highfreq}).

\item \emph{No condensate for any band.} A gauge theorem excludes a Floquet photon condensate for any band structure and to all orders in $\omega_p^{-1}$, including the Kane nonparabolicity~\cite{Kane1957,Kim2026,SqueezingDraft}. For the latter we derive a generalized TRK sum rule that, like the usual one, follows from gauge invariance (Sec.~\ref{sec:gauge}).

\item \emph{Origin of the predicted SRPT: a gauge-invariance violation.} The transition of Ref.~\cite{Wu2026} traces to averaging a Landau-basis matrix element separately from the Hamiltonian. The resulting effective Hamiltonian violates the TRK sum rule, $\Omega_{\rm eff}^2>D\omega_{\rm eff}$, the same gauge-invariance violation behind earlier spurious SRPT predictions. Under driving there is a second, dynamical requirement. Even a model that obeys TRK at every instant is not gauge invariant unless it includes the induced (Faraday) electric field. The time-dependent Hopfield model is accurate for slow modulation without field reversal (Sec.~\ref{sec:highfreq}).

\item \emph{Dynamical Casimir emission.} The dominant resonances are Faraday driven. The induced field pumps pairs of polaritons and low-frequency magnetoplasmons at low threshold. The polariton sum resonance yields two-color photon pairs that are two-mode squeezed (Sec.~\ref{sec:finite}).

\item \emph{Generality.} The rule to modulate and average the minimal-coupling Hamiltonian, not the parameters of its projection onto a basis, applies to any linear, minimally coupled USC system. Magnetic coupling to nonlinear spins is the exception (Sec.~\ref{sec:generality}).
\end{enumerate}

To readers familiar with classical optics, the absence of a condensate in a linear system may seem expected; the point of the present analysis is to make this expectation rigorous for the driven quantum problem, to identify precisely where a model that predicts otherwise departs from it, and to determine what the drive does produce instead.

The rest of the paper is organized as follows. Section~\ref{sec:models} introduces the Hopfield and exact Kohn-mode models and the classical correspondence. Section~\ref{sec:positivity} establishes positivity and proves the Krein-frequency theorem. Section~\ref{sec:gauge} gives the gauge theorem and the Kane Landau ladder. Section~\ref{sec:highfreq} treats the high-frequency limit and the origin of the predicted SRPT. Section~\ref{sec:finite} analyzes finite drive frequencies, loss, emission, and squeezing. Section~\ref{sec:generality} discusses generality, Sec.~\ref{sec:discussion} experimental scales and refinements, and Sec.~\ref{sec:conclusions} concludes. Further technical details are reported in the appendices.

\section{Models}
\label{sec:models}

The electrons and the cavity mode are treated quantum mechanically. The modulating field $B(t)$ and its induced electric field are externally imposed and contain many photons, so we treat them as classical fields. We work in the long-wavelength (local) limit, in which a uniform in-plane cavity field drives transitions only between adjacent Landau levels. For a filled Fermi sea only the transition from the highest occupied level is allowed, and for a parabolic band it carries the entire oscillator strength at the cyclotron frequency (Kohn's theorem). Nonlocal effects, which require a field varying on the scale of the cyclotron orbits, are discussed in Sec.~\ref{sec:discussion}.

\subsection{Hopfield model and its parameter-modulated form}
\label{sec:hopfield}

The static Coulomb-gauge Hamiltonian of Landau polaritons is the bosonic Hopfield model~\cite{Hagenmuller2010,Hopfield1958}
\begin{equation}
\frac{\hat H_{\rm Hop}}{\hbar}=\omega_{\rm cav}\ad\aop+\omega_0\bd\bop+D(\aop+\ad)^2+\Omega(\aop+\ad)(\bop+\bd),
\label{eq:Hop}
\end{equation}
where $\aop$ annihilates a cavity photon of frequency $\omega_{\rm cav}$, and $\bop=N_{\rm LL}^{-1/2}\sum_k\hat c^\dagger_{\nu-1,k}\hat c_{\nu,k}$ annihilates a collective cyclotron excitation~\cite{Hagenmuller2010}. It is bosonic for excitation numbers small compared with the Landau-level degeneracy $N_{\rm LL}$ (for a parabolic band, this restriction is lifted by Kohn's theorem; Sec.~\ref{sec:kohn}), and it is not a two-level system, so there are no saturation effects. The cyclotron frequency is $\omega_0=eB_0/m^*$, with $e$ the elementary charge and $m^*$ the conduction-band effective mass ($0.067\,m_e$ in GaAs). The diamagnetic coefficient is $D=e^2\rho A_0^2S/(2m^*\hbar)=\omega_P^2/4\omega_{\rm cav}$. Here $\rho$ is the sheet density, $S$ the area of the electron gas within the mode, and $A_0=\sqrt{\hbar/2\epsilon_0\epsilon_r\omega_{\rm cav}V}$ the vacuum vector-potential amplitude for a mode volume $V$ and background permittivity $\epsilon_r$. The plasma frequency, $\omega_P^2=e^2\rho S/(\epsilon_0\epsilon_rm^*V)$ for the electron density averaged over the mode, does not depend on $B$. At $B=0$ the dressed cavity frequency is $\sqrt{\omega_{\rm cav}^2+\omega_P^2}$, the classical result for a cavity filled with a Drude plasma.

The static coupling satisfies $\Omega_0^2\equiv\Omega^2=D\omega_0$, and this relation is not an additional assumption. It follows from minimal coupling, i.e., from gauge invariance~\cite{Gustin2023,Garziano2020,Hughes2024,LSA2026}, and it is equivalent to the TRK sum rule. Importantly, in the Coulomb gauge $D$ does not depend on $B$. It is fixed by the electron density, the effective mass, and the mode volume, so when the field changes only $\omega_{\rm cyc}$ and $\Omega$ change, with $\Omega^2/\omega_{\rm cyc}=D$ held fixed. As usual for USC, we characterize the coupling strength by the normalized coupling $\eta\equiv\Omega_0/\omega_{\rm cav}$. We normalize by the cavity frequency rather than by $\omega_0$, because the diamagnetic term remains large even as $\omega_0\to0$, since $D/\omega_{\rm cav}=\eta^2\omega_{\rm cav}/\omega_0$. The USC regime corresponds to $\eta\gtrsim0.1$. Throughout this paper we use $\omega_{\rm cav}=\omega_0$ and $\Omega_0=0.5\,\omega_0$, i.e., $\eta=0.5$, well inside the USC regime and comparable to experiments~\cite{Scalari2012,Bayer2017,Kim2026}. The terms $D(\aop^2+\ad{}^2)$ and $\Omega(\aop\bop+\ad\bd)$ are single- and two-mode squeezing terms; statically they squeeze the ground state, producing virtual photons that are not emitted~\cite{Ciuti2005,SqueezingDraft}.

Under modulation, Ref.~\cite{Wu2026} keeps the Hopfield form, Eq.~\eqref{eq:Hop}, with the operator $\bop$ held fixed, and makes two of its parameters time dependent,
\begin{equation}
\begin{split}
\frac{\hat H_{\rm Hop}(t)}{\hbar}={}&\omega_{\rm cav}\ad\aop+\omega_b(t)\,\bd\bop+D(\aop+\ad)^2\\
&+\Omega(t)(\aop+\ad)(\bop+\bd),
\end{split}
\label{eq:Hopt}
\end{equation}
with $\Omega(t)=\Omega_0\sqrt{|1+\varepsilon\cos\omega_pt|}$ and $D$ unchanged. Here $\omega_b(t)$ is the instantaneous frequency of the matter (cyclotron) excitation in the model, which replaces $\omega_0$ of Eq.~\eqref{eq:Hop}. The absolute value in $\Omega(t)$ appears because $\Omega^2=D|\omega_{\rm cyc}|$. The oscillator strength of the cyclotron transition depends on the magnitude of $B$, while its sign only sets the sense of the cyclotron orbit. We refer to Eq.~\eqref{eq:Hopt} as the \emph{parameter-modulated Hopfield model}.

In Ref.~\cite{Wu2026}, the matter frequency keeps its linear dependence on the field, $\omega_b(t)=\omega_{\rm cyc}(t)=\omega_0[1+\varepsilon\cos\omega_pt]$, whose cycle average is exactly $\omega_0$; we call this the \emph{signed} version. For $\varepsilon>1$, $B(t)$ changes sign during part of each cycle, and $\omega_{\rm cyc}(t)$ then becomes negative, so that the cyclotron excitation acquires a negative energy. Physically, however, the Landau-level spacing is $\hbar|\omega_{\rm cyc}|$ for either sign of $B$; reversing the field reverses the sense of the cyclotron orbit, not the sign of the excitation energy. For comparison, we therefore also consider a version with $\omega_b(t)=|\omega_{\rm cyc}(t)|$, which keeps the excitation energy positive and satisfies $\Omega(t)^2=D\,\omega_b(t)$, i.e., the TRK sum rule, at every instant. Both versions use the same $\Omega(t)$, and they coincide when $B$ does not change sign ($\varepsilon<1$).

\subsection{Exact Kohn-mode model}
\label{sec:kohn}
To go beyond parameter modulation, we include $B(t)$ through its vector potential. For $N$ electrons with parabolic dispersion and pair interactions $\hat V_{ee}$, a uniform field $B(t)\mathbf e_z$ with $\mathbf A_B(\mathbf r,t)=\tfrac12B(t)\,\mathbf e_z\times(\mathbf r-\mathbf r_0)$, where $\mathbf r_0$ is the axis of the modulated flux, and a uniform cavity field $A_0(\aop+\ad)\mathbf e_x$, the Hamiltonian is
\begin{equation}
\begin{split}
\hat H={}&\sum_{i=1}^N\frac{\big[\hat{\mathbf p}_i+e\mathbf A_B(\hat{\mathbf r}_i,t)+eA_0(\aop+\ad)\mathbf e_x\big]^2}{2m^*}\\
&+\hat V_{ee}+\hbar\omega_{\rm cav}\ad\aop .
\end{split}
\label{eq:Hfull}
\end{equation}

The Faraday field is $\mathbf E_F=-\partial_t\mathbf A_B=-\tfrac12\dot B\,\mathbf e_z\times(\mathbf r-\mathbf r_0)$. Since $\mathbf A_B$ is linear in $\mathbf r$, Eq.~\eqref{eq:Hfull} separates exactly into center-of-mass and relative motion. We write $\hat{\mathbf r}_i=\hat{\mathbf R}+\hat{\boldsymbol\xi}_i$ and $\hat{\mathbf p}_i=\hat{\mathbf P}/N+\hat{\boldsymbol\pi}_i$, with $\hat{\mathbf R}=N^{-1}\sum_i\hat{\mathbf r}_i$, $\hat{\mathbf P}=\sum_i\hat{\mathbf p}_i$, and $\sum_i\hat{\boldsymbol\xi}_i=\sum_i\hat{\boldsymbol\pi}_i=0$. Linearity gives $\mathbf A_B(\hat{\mathbf r}_i,t)=\mathbf A_B(\hat{\mathbf R},t)+\tfrac12B(t)\,\mathbf e_z\times\hat{\boldsymbol\xi}_i$, so each kinetic momentum splits as
\begin{equation}
\begin{split}
&\hat{\mathbf p}_i+e\mathbf A_B(\hat{\mathbf r}_i,t)+eA_0(\aop+\ad)\mathbf e_x\\
&\quad=\frac{\hat{\boldsymbol\Pi}}{N}+\Big[\hat{\boldsymbol\pi}_i+\frac{eB(t)}{2}\,\mathbf e_z\times\hat{\boldsymbol\xi}_i\Big],
\end{split}
\end{equation}
where $\hat{\boldsymbol\Pi}$ is the c.m.\ kinetic momentum defined in Eq.~\eqref{eq:HK} below. The bracketed terms sum to zero over $i$, so the cross terms cancel when the squares are summed, and Eq.~\eqref{eq:Hfull} becomes $\hat H=\hat H_{\rm c.m.}+\hat H_{\rm rel}$ with
\begin{align}
\hat H_{\rm c.m.}&=\frac{\hat{\boldsymbol\Pi}^2}{2M}+\hbar\omega_{\rm cav}\ad\aop ,\\
\hat H_{\rm rel}&=\sum_{i=1}^N\frac{1}{2m^*}\Big[\hat{\boldsymbol\pi}_i+\frac{eB(t)}{2}\,\mathbf e_z\times\hat{\boldsymbol\xi}_i\Big]^2+\hat V_{ee}.
\end{align}

The relative part contains neither $\hat{\mathbf R}$ nor $\aop$, and $\hat V_{ee}$ depends only on the relative coordinates~\cite{Kohn1961,Brey1989,MaksymChakraborty1990,Dobson1994,Rokaj2023}. Since the c.m.\ and relative operators commute, the two parts evolve independently. The relative motion feels $B(t)$, including its own Faraday field about the c.m., but it does not couple to the cavity, and it does not depend on the flux axis $\mathbf r_0$. The reduction is exact for interacting electrons at any filling.
Thus, \emph{the c.m.\ ladder operator defined below is exactly bosonic, so no Holstein--Primakoff or dilute-excitation approximation is involved}, and Pauli blocking and filling affect only the relative motion. The separation holds for any filling factor, any $B(t)$, and through $B=0$. The cavity--matter problem is therefore exactly
\begin{equation}
\begin{split}
\hat H_K(t)&=\frac{\hat{\boldsymbol\Pi}^2}{2M}+\frac{M\omega_d^2}{2}|\hat{\mathbf R}-\mathbf r_0|^2+\hbar\omega_{\rm cav}\ad\aop,\\
\hat{\boldsymbol\Pi}&=\hat{\mathbf P}-Q\mathbf A_B(\hat{\mathbf R},t)-QA_0(\aop+\ad)\,\mathbf e_x ,
\end{split}
\label{eq:HK}
\end{equation}
where the subscript $K$ stands for the Kohn (c.m.) mode, $Q=-Ne$ is the total charge, $M=Nm^*$ the total mass, and $\hat{\boldsymbol\Pi}$ the kinetic momentum of the c.m. The second term is a harmonic restoring force on the c.m., characterized by the frequency $\omega_d$. Its origin is discussed below; it acts on the c.m.\ alone and does not spoil the separation.

Although $B$ is classical, $\mathbf A_B(\hat{\mathbf R},t)$ is an operator, since it depends on the c.m.\ position. Writing $\hat{\boldsymbol\Pi}_B=\hat{\mathbf P}-Q\mathbf A_B(\hat{\mathbf R})$ for its field part, one finds $[\hat\Pi_{B,x},\hat\Pi_{B,y}]=i\hbar QB$, which produces the Landau ladder. For static $B$, with $s={\rm sgn}(QB)$, the operator $\bop=(\hat\Pi_{B,x}+is\hat\Pi_{B,y})/\sqrt{2\hbar|QB|}$ satisfies $[\bop,\bd]=1$, with $\hat{\boldsymbol\Pi}_B^2/2M=\hbar|\omega_{\rm cyc}|(\bd\bop+\tfrac12)$ and $\hat\Pi_{B,x}=\sqrt{\hbar|QB|/2}\,(\bop+\bd)$, where $|\omega_{\rm cyc}|=|QB|/M$. Expanding $\hat{\boldsymbol\Pi}^2/2M$ then gives
\begin{equation}
\begin{split}
\frac{\hat{\boldsymbol\Pi}^2}{2M}={}&\hbar|\omega_{\rm cyc}|\big(\bd\bop+\tfrac12\big)-\frac{QA_0}{M}\sqrt{\frac{\hbar|QB|}{2}}\,(\aop+\ad)(\bop+\bd)\\
&+\frac{Q^2A_0^2}{2M}(\aop+\ad)^2 ,
\end{split}
\end{equation}
which is Eq.~\eqref{eq:Hop} with $\hbar D=Q^2A_0^2/2M$ and $\Omega^2=D|\omega_{\rm cyc}|$. The Kohn ladder is exactly harmonic. For $B<0$ the roles of $\hat\Pi_{B,x}\pm i\hat\Pi_{B,y}$ are exchanged; this is the origin of the sign ambiguity in parameter-modulated models. For a parabolic band, $\bop$ acting on the filled Fermi sea at integer filling is, up to normalization, the collective operator of Ref.~\cite{Hagenmuller2010}, since all other $n\to n+1$ terms are Pauli blocked.

The guiding center $\hat{\mathbf R}_{\rm gc}=\hat{\mathbf R}-\mathbf r_0-\mathbf e_z\times\hat{\boldsymbol\Pi}_B/QB$ is the center of the cyclotron orbit [Fig.~\ref{fig:schematic}(b)]. The c.m.\ position is the sum of this center and the cyclotron radius vector $\mathbf e_z\times\hat{\boldsymbol\Pi}_B/QB$, which rotates at the cyclotron frequency. In a static, uniform field the guiding center stays fixed, and it moves only under additional forces, such as the edge restoring force or the Faraday field, which make it drift. Formally, $\hat{\mathbf R}_{\rm gc}$ commutes with $\hat{\boldsymbol\Pi}_B$ and does not appear in $\hat{\boldsymbol\Pi}^2/2M$. For an infinite sample it is a zero mode, since a uniform translation of the electron gas costs no energy.

Every real sample is finite, and so is the modulated flux. Displacing the electron gas then builds up charge at its edges, whose depolarization field restores the c.m. We model this by the harmonic term in Eq.~\eqref{eq:HK}, where $\omega_d$ is the frequency at which the electron gas oscillates as a whole at $B=0$ (its dipolar plasmon). The static matter modes are then the Fock--Darwin modes, $\omega_\pm^{\rm FD}=\sqrt{\omega_d^2+\omega_0^2/4}\pm\omega_0/2$~\cite{Fock1928,Darwin1931}. The lower one is the low (edge) magnetoplasmon, $\omega_{\rm mp}\simeq\omega_d^2/\omega_0$, and the upper one mixes with the cavity to form the polaritons $\omega_\pm$. For $\omega_d\to0$ the magnetoplasmon becomes the guiding-center zero mode. The finite-sample model is the physical one, and $\omega_d=0$ is a singular limit.

\subsection{The modulated problem in the static Landau basis}
\label{sec:staticbasis}
Writing Eq.~\eqref{eq:HK} in the static Landau basis (that of $B_0$) shows precisely what the parameter-modulated model omits. Let $\hat{\boldsymbol\Pi}_0$ and $\hat{\mathbf R}_{\rm gc}$ refer to $B_0$, and write $B(t)=B_0+\delta B(t)$.
The vector potential $\mathbf A_B$ is linear in $B$, so that $\hat{\boldsymbol\Pi}_B(t)=\hat{\boldsymbol\Pi}_0-\tfrac12Q\,\delta B\,\mathbf e_z\times(\hat{\mathbf R}-\mathbf r_0)$. Writing the c.m.\ position as guiding center plus cyclotron radius, $\hat{\mathbf R}-\mathbf r_0=\hat{\mathbf R}_{\rm gc}+\mathbf e_z\times\hat{\boldsymbol\Pi}_0/QB_0$, and using $\mathbf e_z\times(\mathbf e_z\times\mathbf v)=-\mathbf v$ for in-plane vectors, gives the exact identity:
\begin{equation}
\hat{\boldsymbol\Pi}_B(t)=\Big(1+\frac{\delta B}{2B_0}\Big)\hat{\boldsymbol\Pi}_0-\frac{Q\delta B}{2}\,\mathbf e_z\times\hat{\mathbf R}_{\rm gc}
\label{eq:Pidecomp}
\end{equation}
and therefore
\begin{equation}
\begin{split}
\frac{\hat{\boldsymbol\Pi}_B^2}{2M}={}&\Big(1+\frac{\delta B}{2B_0}\Big)^2\frac{\hat{\boldsymbol\Pi}_0^2}{2M}+\frac{(Q\delta B)^2}{8M}|\hat{\mathbf R}_{\rm gc}|^2\\
&-\Big(1+\frac{\delta B}{2B_0}\Big)\frac{Q\delta B}{2M}\,\hat{\boldsymbol\Pi}_0\cdot(\mathbf e_z\times\hat{\mathbf R}_{\rm gc}).
\end{split}
\label{eq:static_basis}
\end{equation}

The cavity coupling similarly becomes $-(QA_0/M)(\aop+\ad)[(1+\delta B/2B_0)\hat\Pi_{0,x}-\tfrac12Q\delta B(\mathbf e_z\times\hat{\mathbf R}_{\rm gc})_x]$. Comparing with the parameter-modulated model:
\begin{itemize}
\item \emph{First order in $\varepsilon$.} The cyclotron term, $\omega_0(1+\delta B/2B_0)^2\simeq\omega_{\rm cyc}(t)$, and the coupling, $\Omega_0(1+\delta B/2B_0)\simeq\Omega_0\sqrt{1+\delta B/B_0}$, agree. What the parameter-modulated model lacks are the guiding-center terms, i.e., the cyclotron--guiding-center and cavity--guiding-center couplings. These are the Faraday field in this basis. They contain pair-creation terms such as $\bd\hat c_{\rm gc}^\dagger$ and generate the \emph{Faraday tongues} of Sec.~\ref{sec:finite}.
\item \emph{Second order in $\varepsilon$.} The cyclotron coefficient is $\omega_0(1+\delta B/2B_0)^2\ge0$, rather than $\omega_0(1+\delta B/B_0)$ (signed) or $\omega_0|1+\delta B/B_0|$. The coupling is linear in $\delta B$, not a square root, and the guiding center feels the ponderomotive trap $(Q\delta B)^2|\hat{\mathbf R}_{\rm gc}|^2/8M$. No sign prescription is needed when $B$ reverses.
\end{itemize}
The static Landau levels, together with the guiding center, therefore remain a complete and perfectly good basis. What the parameter-modulated model omits are terms of the Hamiltonian, not states of the basis. Equivalently, one can use instantaneous ladder operators $\bop_t$, built from $\hat{\boldsymbol\Pi}_{B(t)}$ in the same way as $\bop$ is built from $\hat{\boldsymbol\Pi}_{B_0}$. 
These are Schr\"odinger-picture operators with an explicit, parametric time dependence through $B(t)$, not Heisenberg-picture operators. They are related to the static ones by a time-dependent unitary transformation, $\bop_t=\hat U_t\bop\hat U_t^\dagger$, i.e., a time-dependent change of basis to the instantaneous Landau levels. 

In this instantaneous basis, the Hamiltonian at each fixed time $t$ has exactly the Hopfield form of Eq.~\eqref{eq:Hopt}, with $\bop$ replaced by $\bop_t$, $\omega_b=|\omega_{\rm cyc}(t)|$, and $\Omega\propto\sqrt{|B(t)|}$. The basis itself, however, changes with time. Writing the state in this basis, $|\tilde\psi\rangle=\hat U_t^\dagger|\psi\rangle$, the Schr\"odinger equation is generated by
\begin{equation}
\hat{\tilde H}(t)=\hat U_t^\dagger\hat H\hat U_t-i\hbar\,\hat U_t^\dagger\dot{\hat U}_t .
\label{eq:Hinst}
\end{equation}
The first term is the instantaneous Hopfield Hamiltonian, which is all that the parameter-modulated model retains. The second term arises because the basis moves; it is proportional to $\dot B$, couples $\bop_t$ to $\bd_t$ and to the guiding center, and is the Faraday field expressed in the instantaneous basis. Relative to the cyclotron frequency, it is of order $|\dot B/B|/\omega_0\sim\varepsilon\omega_p/\omega_0$. It can therefore be neglected, and the parameter-modulated model is accurate, when the modulation is slow ($\varepsilon\omega_p\ll\omega_0$) and $B$ does not change sign. However, if $B$ passes through zero, the instantaneous Landau basis is undefined and $\dot B/B$ diverges, so no such approximation holds.

\subsection{Classical correspondence and the quantum content}
\label{sec:classical}
Equation~\eqref{eq:HK} is quadratic, so its  Heisenberg equations coincide with the classical Newton--Maxwell equations,
\begin{align}
M\ddot{\mathbf R}&=Q\big[-\tfrac12\dot B\,\mathbf e_z\times(\mathbf R-\mathbf r_0)+B\,\dot{\mathbf R}\times\mathbf e_z\big]\nonumber\\
&\quad-Q\dot{\mathcal A}\,\mathbf e_x-M\omega_d^2(\mathbf R-\mathbf r_0),\label{eq:newton}\\
\ddot{\mathcal A}+\omega_{\rm cav}^2\mathcal A&=Q\dot R_x/\epsilon_0\epsilon_rV ,\label{eq:maxwell}
\end{align}
with $\mathcal A=A_0(\aop+\ad)$. These are the Lorentz-force equation for a charge $Q$ in $B(t)$, including its Faraday field, and the mode equation with the cavity driven by the total current $Q\dot R_x$. 

Note that 
the $A^2$ term does not appear explicitly in these equations, but it is not missing. In the Hamiltonian, it arises from expanding $(\hat{\mathbf P}-Q\hat{\mathbf A})^2/2M$; in the equations of motion, the same physics is contained in the relation between the canonical momentum and the velocity, $M\dot{\mathbf R}=\mathbf P-Q\mathbf A$. The current that drives the cavity, $Q\dot R_x$, is therefore the full gauge-invariant current, including its diamagnetic part $-Q^2A/M$.
Newton's equations, written in terms of the physical fields $\mathbf E$ and $\mathbf B$ and the velocity, contain no gauge choice and no truncation, so they automatically satisfy the TRK sum rule, since at high frequency the charges respond as free particles of mass $M$. This is why Maxwell--Lorentz solvers are automatically gauge consistent~\cite{PyCharge}, and why the gauge problems of USC arise only when the matter is truncated to a few levels or a model is written in a fixed basis.

The static Hopfield model, Eq.~\eqref{eq:Hop}, likewise maps onto two position-coupled oscillators~\cite{ChangPRA2025},
\begin{equation}
\begin{split}
\ddot x_a+\omega_{\rm cav}(\omega_{\rm cav}+4D)\,x_a&=-2\omega_{\rm cav}\Omega\,x_b,\\
\ddot x_b+\omega_b^2x_b&=-2\omega_b\Omega\,x_a ,
\end{split}
\end{equation}
where the renormalized cavity frequency keeps the system stable. With TRK, $\omega_{\rm cav}(\omega_{\rm cav}+4D)\omega_b^2-4\omega_{\rm cav}\omega_b\Omega^2=\omega_{\rm cav}^2\omega_b^2>0$. These second-order equations are obtained by eliminating the momenta from Hamilton's (first-order) equations, e.g., $\dot x_b=\omega_bp_b$. If $\omega_b$ depends on time, differentiating again gives $\ddot x_b=\omega_b\dot p_b+\dot\omega_bp_b$, which contains an extra term $(\dot\omega_b/\omega_b)\dot x_b$. Simply replacing $\omega_b\to\omega_b(t)$ in the second-order equations is therefore not equivalent to modulating the Hamiltonian. A physical modulation acts on the Hamiltonian, so it is the Hamiltonian, or equivalently the first-order equations, that must be modulated, with the second-order form derived afterward.

Since the Heisenberg equations of a quadratic Hamiltonian coincide with the classical equations of motion, the stability, quasienergies, and resonance conditions are classical properties. Quantum mechanics enters in three places: (i)~the emission of pairs from the vacuum, since a noiseless classical system stays at rest; (ii)~squeezing below the vacuum level; and (iii)~the quantum correlations (entanglement) between the emitted photons at $\omega_-$ and $\omega_+$. Unlike light--matter entanglement inside the cavity, which depends on how the subsystems are defined and hence on the gauge, these correlations refer to the output fields outside the cavity and are directly measurable.
For a prescribed classical modulation the Hamiltonian stays quadratic, so the DCE here is linear parametric amplification of vacuum fluctuations. Genuinely nonclassical dynamics beyond Gaussian states would require nonlinearity, such as the Kane nonparabolicity, which turns the dressed mode into a weak Kerr oscillator~\cite{SqueezingDraft}, or back-action on a quantized modulator.

\section{Positivity and stability to all orders}
\label{sec:positivity}

This section contains the central stability result, and we first summarize it in words. The exact Hamiltonian is a sum of squares at every instant, so its energy can never run away statically, and no static photon field can lower it. This alone does not settle the driven dynamics, since a periodic drive can still pump energy into the system. Krein's theorem settles this question.
If the drive is faster than twice the system's highest instantaneous mode frequency, averaged over a cycle, then no pair of modes can be excited resonantly, and the system remains stable however long it is driven, to all orders in the drive. Below that frequency, the only way stability can be lost is the resonant creation of pairs of excitations, i.e., the dynamical Casimir effect. Subsections~\ref{sec:floquetnotation} and~\ref{sec:instpos} set up the notation and establish positivity, Sec.~\ref{sec:kreinthm} states and proves the theorem, and Sec.~\ref{sec:belowK} describes what happens below the Krein frequency.

\subsection{Quadratic Hamiltonians and Floquet multipliers}
\label{sec:floquetnotation}
We write all models as $\hat H(t)=\tfrac12\hat X^{\rm T}\mathcal H(t)\hat X$, where $\hat X=(\hat q_1,\dots,\hat q_n,\hat p_1,\dots,\hat p_n)^{\rm T}$ collects canonical quadratures, $[\hat X_j,\hat X_k]=iJ_{jk}$ with $\hbar=1$ and $J=\big(\begin{smallmatrix}0&\openone\\-\openone&0\end{smallmatrix}\big)$, and $\mathcal H(t)$ is a real symmetric $T$-periodic matrix, $T=2\pi/\omega_p$. For a boson system,  $\hat q=(\aop+\ad)/\sqrt2$ and $\hat p=(\aop-\ad)/i\sqrt2$. 

The Heisenberg equations, $\dot{\hat X}=J\mathcal H(t)\hat X$, have the same form as Hamilton's equations, so the operators evolve as $\hat X(t)=\Phi(t)\hat X(0)$ with the classical propagator $\Phi(t)$. Expectation values therefore follow the classical trajectories exactly, and the covariance matrix evolves as $\Sigma(t)=\Phi(t)\Sigma(0)\Phi(t)^{\rm T}$. Quantum mechanics enters only through the initial state, e.g., the vacuum fluctuations $\Sigma(0)=\tfrac12\openone$, and not through the dynamics.

The monodromy matrix $\Phi(T)$ propagates $X$ over one period.\footnote{The monodromy matrix is the one-period propagator of the classical (or Heisenberg) equations of motion, $X(T)=\Phi(T)X(0)$. For a periodically driven system it plays the role of the Floquet operator. After $n$ periods, $X(nT)=\Phi(T)^nX(0)$, so its eigenvalues determine whether solutions stay bounded (all $|\mu|=1$) or grow ($|\mu|>1$).} Its eigenvalues $\mu$ are the Floquet multipliers, $\lambda=\max\ln|\mu|/T$ is the Floquet exponent (the growth rate of the fastest-growing solution), and $\arg\mu/T$ (mod $\omega_p$) are the quasienergies. We write $\mathcal H\succeq0$ ($\mathcal H\succ0$) if all eigenvalues of $\mathcal H$ are nonnegative (positive), i.e., if the classical energy is bounded below (has a unique minimum), and $\langle\cdot\rangle=T^{-1}\int_0^T(\cdot)\,dt$ for time averages. For a multiplier on the unit circle with eigenvector $v$, the Krein signature, the sign of $iv^\dagger Jv$, distinguishes positive-energy oscillations (first kind, the ordinary quanta $\hat c_k$) from their negative-energy conjugates (second kind).

\subsection{Instantaneous positivity}
\label{sec:instpos}
For Eq.~\eqref{eq:HK}, $\mathcal H(t)\succeq0$ at every instant, because the Hamiltonian is a sum of squares. For the Hopfield model, positivity is decided by a Schur complement~\cite{ZhangSchur2005,HornJohnson2013}. For a symmetric block matrix $V=\big(\begin{smallmatrix}A&B\\B^{\rm T}&C\end{smallmatrix}\big)$ with $C\succ0$, $V\succeq0$ if and only if $A-BC^{-1}B^{\rm T}\succeq0$. The Hopfield potential block has $A=\omega_{\rm cav}+4D$, $B=2\Omega$, and $C=\omega_b$, and the kinetic block is ${\rm diag}(\omega_{\rm cav},\omega_b)$. The Schur complement, $\omega_{\rm cav}+4(D-\Omega^2/\omega_b)$, is the effective spring constant of the cavity field for a slow (static) displacement, once the matter has adjusted to it. If it became negative, a static photon field would lower the energy, which is the signature of an SRPT. Positivity is therefore equivalent to the instantaneous TRK inequality
\begin{equation}
D\ge\Omega(t)^2/\omega_b(t),
\label{eq:TRKinst}
\end{equation}
and the TRK sum rule makes the inequality an equality, so  the paramagnetic and diamagnetic static responses to a uniform $A$ cancel, as required for a pure gauge~\cite{Andolina2019}. 

With $\omega_b(t)=|\omega_{\rm cyc}(t)|$, Eq.~\eqref{eq:TRKinst} holds at all times. With the signed $\omega_{\rm cyc}(t)$ of Ref.~\cite{Wu2026}, $C=\omega_b<0$ whenever $B(t)<0$, and $\mathcal H(t)$ is indefinite. During that part of each cycle, the instantaneous energy is unbounded below along some directions in phase space, i.e., the model has a negative-energy mode and violates the instantaneous TRK inequality. Positivity, and with it the stability guarantees below, are then lost. This is the origin of the instability found in Ref.~\cite{Wu2026}.

Positivity has an immediate consequence at lowest order of the Floquet--Magnus expansion~\cite{Eckardt2015,Bukov2015}. The effective Hamiltonian $\tfrac12\hat X^{\rm T}\langle \mathcal H\rangle\hat X$ has $\langle \mathcal H\rangle$ equal to an average of positive matrices, which is positive, \emph{so no SRPT occurs}. 
For the Hopfield model with $|\omega_{\rm cyc}|$, for example, averaging the parameters gives $\Omega_{\rm eff}^2=D\langle\sqrt{|\omega_{\rm cyc}|}\rangle^2\le D\langle|\omega_{\rm cyc}|\rangle=D\,\omega_{\rm eff}$, so the averaged model still satisfies the TRK inequality.
 The statement extends to modulation of any parameter that respects Eq.~\eqref{eq:TRKinst}.

\subsection{The Krein-frequency theorem}
\label{sec:kreinthm}
The statement above holds only at lowest order of the high-frequency expansion. Krein's theory of periodically driven linear Hamiltonian systems of positive type~\cite{Krein1955,GelfandLidskii1955,YakubovichStarzhinskii1975,Ekeland1990} extends it to all orders of that expansion. Thus, it establishes stability of the exact driven dynamics for every drive frequency above a computable threshold, not only in the limit $\omega_p\to\infty$.

\begin{theorem}[Krein frequency]
\label{thm:krein}
Let $\mathcal H(t)\succeq0$ be $T$-periodic with $\langle \mathcal H\rangle\succ0$. Then $\dot X=J\mathcal H(t)X$ is strongly stable, i.e., stable together with all sufficiently small Hamiltonian perturbations, whenever
\begin{equation}
\omega_p>\omega_K(\varepsilon)\equiv2\min_S\big\langle\lambda_{\max}[S^{\rm T}\mathcal H(t)S]\big\rangle ,
\label{eq:wK}
\end{equation}
where $\lambda_{\max}$ is the largest eigenvalue and $S$ ranges over constant symplectic matrices.
\end{theorem}

In words, the right-hand side is twice the cycle-averaged highest frequency of the system. At each instant $t$, the largest eigenvalue $\lambda_{\max}$ of the Hamiltonian matrix $\mathcal H(t)$ is an upper bound on the frequencies at which the system can oscillate at that moment; it equals the highest mode frequency in variables where each oscillator's coordinate and momentum enter symmetrically, as for $\tfrac12\omega(q^2+p^2)$. The minimization over constant symplectic transformations $S$, i.e., time-independent canonical changes of variables such as rescaling a coordinate by $s$ and its momentum by $1/s$, finds the variables that make this bound as tight as possible; it does not change the dynamics. The time average $\langle\cdot\rangle$ over one drive period then gives the typical highest mode frequency during the cycle, and the factor of 2 reflects that instabilities create excitations in pairs. Equation~\eqref{eq:wK} therefore states that a drive faster than twice the cycle-averaged highest mode frequency is too fast to resonantly create any pair of excitations, and the system is then stable to all orders.

\emph{Physical meaning.} In the absence of driving, the highest pair resonance is at $2\omega_+$, and indeed Eq.~\eqref{eq:wK} gives $\omega_K\to2\omega_+$ as $\varepsilon\to0$, so the bound is sharp. For deep modulation, $\omega_K\simeq2\langle|\omega_{\rm cyc}|\rangle$. The criterion requires only the instantaneous Hamiltonian, not a solution of the dynamics, and it holds for any waveform and depth, including field reversal. Formally, $X=SY$ gives $\dot Y=JS^{\rm T}\mathcal HSY$, so the dynamics, and hence the stability, are unchanged, while $\mathcal H\to S^{\rm T}\mathcal HS$; how we optimize over $S$ is described in Appendix~\ref{app:numerics}.

\emph{Proof.} 
Step~1 rescales time by the instantaneous mode frequency, so that the drive frequency condition becomes a condition on the period. Step~2 uses an elementary inequality to show that, under this condition, no solution can change sign every period, which is how instabilities first appear for modes of opposite energy sign. Step~3 uses Krein's observation that modes of positive and negative energy rotate in opposite directions around the unit circle as the drive is turned up, so they can never meet. The three steps are as follows.

(1)~\emph{Reparametrization.} Set $d\tau=\lambda_{\max}(t)\,dt$, i.e., measure time in units of the instantaneous highest frequency. This changes only the rate at which a trajectory is traversed, not the trajectory itself, so the one-period map $X(0)\mapsto X(T)$ is unchanged. (If $\lambda_{\max}$ vanishes on an interval, then $\mathcal H=0$ there and nothing evolves, so that interval can be skipped.) In the new time variable, $dX/d\tau=J\tilde{\mathcal H}(\tau)X$ with $0\preceq\tilde{\mathcal H}=\mathcal H/\lambda_{\max}\preceq\openone$, and the period becomes $\tau_T=\int_0^T\lambda_{\max}\,dt=T\langle\lambda_{\max}\rangle$. Since $T=2\pi/\omega_p$ and, for $S=\openone$, $\omega_K=2\langle\lambda_{\max}\rangle$, the condition $\omega_p>\omega_K$ is equivalent to $\tau_T<\pi$.

(2)~\emph{No multiplier reaches $-1$.} Embed the system in the family $dX/d\tau=sJ\tilde{\mathcal H}X$ with $s\in(0,1]$. A multiplier $-1$ means an antiperiodic solution, $X(\tau+\tau_T)=-X(\tau)$, whose Fourier components have frequencies $(2k+1)\pi/\tau_T$. Since $|c^\dagger Jc|\le|c|^2$ for any vector $c$, a 
Wirtinger-type inequality\footnote{Wirtinger's inequality bounds a function by its derivative. For a $\tau_T$-periodic function with zero mean, $\int_0^{\tau_T}|f|^2\le(\tau_T/2\pi)^2\int_0^{\tau_T}|\dot f|^2$~\cite{HardyLittlewoodPolya1952}. The form used here follows the same idea. An antiperiodic function contains only the Fourier frequencies $\omega_k=(2k+1)\pi/\tau_T$, all with $|\omega_k|\ge\pi/\tau_T$. Writing $X=\sum_kc_ke^{i\omega_k\tau}$, one has $\int X^\dagger J\dot X=i\tau_T\sum_k\omega_kc_k^\dagger Jc_k$ and $\int|\dot X|^2=\tau_T\sum_k\omega_k^2|c_k|^2$. Since $|c_k^\dagger Jc_k|\le|c_k|^2$ and $|\omega_k|\le\omega_k^2\tau_T/\pi$, the inequality follows.} follows:
\begin{equation}
\Big|\int_0^{\tau_T}X^\dagger J\dot X\,d\tau\Big|\le\frac{\tau_T}{\pi}\int_0^{\tau_T}|\dot X|^2d\tau .
\end{equation}
The equation of motion gives $J\dot X=-s\tilde{\mathcal H}X$, so the left side equals $s\int X^\dagger\tilde{\mathcal H}X\,d\tau$. Also $|\dot X|^2=s^2|\tilde{\mathcal H}X|^2\le s^2X^\dagger\tilde{\mathcal H}X$, because $\tilde{\mathcal H}^2\preceq\tilde{\mathcal H}$. Hence $s\ge\pi/\tau_T$, unless $\tilde{\mathcal H}X\equiv0$, in which case $X$ is constant and antiperiodic, i.e., zero. For $\tau_T<\pi$, no multiplier reaches $-1$ for any $s\in(0,1]$.

(3)~\emph{Krein monotonicity.} As $s\to0^+$, the monodromy is $\exp(s\tau_TJ\langle\tilde{\mathcal H}\rangle)$. For $\langle\tilde{\mathcal H}\rangle\succ0$ its multipliers $e^{\pm is\tau_T\omega_k}$ split into first kind (upper half circle) and second kind (lower half circle). For a simple multiplier $\mu=e^{i\theta}$ with eigenvector $v$ and Floquet solution $X(\tau)=\Phi(\tau)v$, the left eigenvector of the real symplectic monodromy is $v^\dagger J$; this follows from $\Phi^{\rm T}J\Phi=J$ and $|\mu|=1$. Variation of constants, $\partial_s\Phi(\tau_T)=\Phi(\tau_T)\int_0^{\tau_T}\Phi^{-1}J\tilde{\mathcal H}\Phi\,d\tau$, together with $v^\dagger J\Phi(\tau)^{-1}=X(\tau)^\dagger J$, gives
\begin{equation}
\frac{d\theta}{ds}=\frac{\int_0^{\tau_T}X^\dagger\tilde{\mathcal H}X\,d\tau}{-i\,v^\dagger Jv}.
\label{eq:kreinmono}
\end{equation}
The numerator is positive. It vanishes only if $\tilde{\mathcal H}X\equiv0$, i.e., for a constant $X$ with $\langle\tilde{\mathcal H}\rangle X=0$, which is excluded. The denominator is real and nonzero, and its sign is the Krein signature, which is conserved along the solution. First-kind multipliers therefore rotate strictly counterclockwise and second-kind ones clockwise. 

If several multipliers of the same kind coincide (degenerate multipliers, i.e., several independent Floquet solutions with the same multiplier), the same argument applies to the cluster as a whole, whose total phase still increases monotonically~\cite{Krein1955,GelfandLidskii1955,Ekeland1990}. A first-kind multiplier can leave the upper half circle only through $-1$, which step~2 excludes. Multipliers of opposite kind therefore never meet, and by the Krein--Gel'fand--Lidskii theorem~\cite{Krein1955,GelfandLidskii1955,Moser1958} the system is strongly stable. Applying the argument in the variables $Y=S^{-1}X$ gives the $\min_S$. This completes the proof.

Step~2 needs only an elementary inequality for antiperiodic functions, and step~3 is Krein's classical monotonicity, derived here in self-contained form. Without step~1, the same argument applied directly in $t$ would require $T\sup_t\lambda_{\max}<\pi$. Since the maximum of $\lambda_{\max}$ over a cycle exceeds its average, this condition is more restrictive, especially for strongly pulsed drives. The reparametrization replaces the maximum by the time average.

\begin{figure*}[t]
\centering\includegraphics[width=\textwidth]{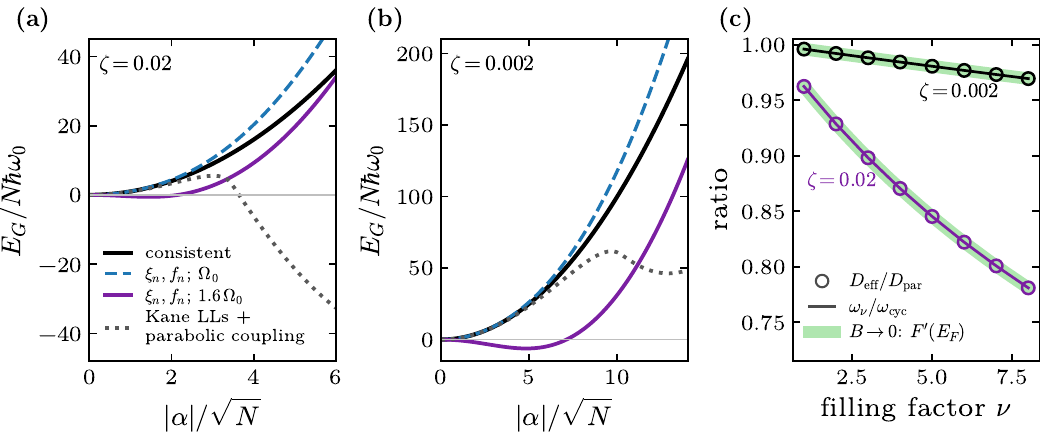}
\caption{Kane Landau ladder. (a,b)~Mean-field energy per electron for $\zeta=\hbar\omega_{\rm cyc}/E_g^*=0.02$ and 0.002. Black: gauge consistent. Blue dashed and purple: model with nonparabolic energies $\xi_n$ and matrix elements $f_n$ (Ref.~\cite{Wu2026}) at $\Omega_{\rm eff}=\Omega_0$ and $1.6\,\Omega_0$. Dotted: Kane Landau-level energies with the parabolic coupling and $D$ added afterwards, i.e., the band is projected before the light--matter coupling is introduced, which violates gauge invariance. (c)~Generalized TRK sum rule, Eq.~\eqref{eq:genTRK}: $D_{\rm eff}/D_{\rm par}$ (circles) equals $\omega_\nu/\omega_{\rm cyc}$ (lines); green band: $B\to0$ limit $F'(E_F)$; black: $\zeta=0.002$, purple: $\zeta=0.02$.}
\label{fig:kane}
\end{figure*}

\subsection{Structure below the Krein frequency}
\label{sec:belowK}
Below $\omega_K$, stability can be lost only through collisions of multipliers of opposite Krein signature~\cite{GelfandLidskii1955,Moser1958,Arnold1989}, i.e., at sum resonances $\tilde\epsilon_i+\tilde\epsilon_j=n\omega_p$ of the dressed quasienergies. These are pair creation, the DCE. They appear as instability (Arnold) tongues, namely wedge-shaped regions of the $(\varepsilon,\omega_p)$ plane that stem from the resonance points at $\varepsilon=0$ and widen as $\varepsilon^n$, as in the Ince--Strutt chart of the Mathieu equation~\cite{LandauMechanics,Kovacic2018}. Difference resonances give avoided crossings, never instabilities.

A first-order estimate of the tongue rates follows from the rotating-wave approximation. Let $\hat c_k$ be the normal modes of $\langle \mathcal H\rangle$. The part of $\tfrac12\hat X^{\rm T}[\mathcal H(t)-\langle \mathcal H\rangle]\hat X$ that creates pairs is $\tfrac12\sum_{ij}K_{ij}(t)\hat c_i^\dagger\hat c_j^\dagger+{\rm H.c.}$ Near $\omega_p\simeq\omega_i+\omega_j$ this gives a peak rate
\begin{equation}
\lambda_{ij}\simeq|K^{(1)}_{ij}|,\qquad K^{(1)}=\langle K(t)e^{i\omega_pt}\rangle ,
\label{eq:rates}
\end{equation}
which reproduces the exact $n=1$ peak rates to within about 10\% (Appendix~\ref{app:numerics}).

\section{No condensate for any band or drive}
\label{sec:gauge}

\subsection{Gauge theorem}
Positivity governs the dynamics of fluctuations. Whether a static photon field can lower the energy is settled by a simpler argument, which also covers band nonparabolicity. Take the most general single-band electronic Hamiltonian,
\begin{equation}
\hat H_{\rm el}(t;\mathbf A_c)=\sum_jE\big(\hat{\mathbf p}_j+e\mathbf A_B(\hat{\mathbf r}_j,t)+e\mathbf A_c\big)+\hat V_{ee},
\end{equation}
where $E(\mathbf p)$ is the band dispersion with minimal coupling applied before projection (Peierls substitution; equivalently the projected (gauge-invariant) unitary of Refs.~\cite{DiStefano2019,Garziano2020,Gustin2023,Kim2026}), and $\mathbf A_c=A_0(\alpha+\alpha^*)\mathbf e_x$ is the mean cavity field. 
The time-independent unitary $\hat U=\exp(-ie\mathbf A_c\cdot\sum_j\hat{\mathbf r}_j/\hbar)$ gives
\begin{equation}
\hat H_{\rm el}(t;\mathbf A_c)=\hat U\hat H_{\rm el}(t;0)\hat U^\dagger
\end{equation}
at every instant, because $\mathbf A_B$ and $\hat V_{ee}$ depend only on positions. 

The Floquet operator, its quasienergies, and the Floquet--Magnus Hamiltonian at every order are therefore independent of $\alpha$. The mean-field quasienergy is then $E_{\rm MF}(\alpha)=\hbar\omega_{\rm cav}|\alpha|^2+{\rm const}$, which has no minimum at $\alpha\neq0$. This extends the equilibrium no-go theorem~\cite{Andolina2019,Andolina2022} to periodic driving, to any band structure, and to all orders in $\omega_p^{-1}$. It is the relaxed (static) energy relevant to a condensate. The unrelaxed $B=0$ Kane potential of Ref.~\cite{SqueezingDraft}, at fixed canonical momenta and relevant to the fast cavity dynamics, is monotonic, so neither level of description yields a minimum. Reference~\cite{Wu2026} noted that an equally spaced Landau-level ladder gives no condensate, and obtained finite-field minima only after adding band nonparabolicity. 

The gauge theorem shows that the absence of a condensate persists for any band structure, provided the coupling is introduced by minimal substitution. Approximate models violate the theorem when the coupling is not obtained by minimal substitution in the full electronic Hamiltonian, as we now show.

\subsection{Kane Landau ladder and a generalized TRK sum rule}
\label{sec:kane}
In the two-band Kane model, $E(1+E/E_g^*)=\hat{\boldsymbol\Pi}^2/2m^*$ on the conduction band~\cite{Kane1957}, so $E=F(\hat{\boldsymbol\Pi}^2/2m^*)$ with
\begin{equation}
F(u)=\tfrac12E_g^*\big[\sqrt{1+4u/E_g^*}-1\big],
\end{equation}
the function behind the all-orders Kane potential of Ref.~\cite{SqueezingDraft} (spin, Zeeman, and spin--orbit terms neglected). The Landau levels, $E_n=F(\hbar\omega_{\rm cyc}(n+\tfrac12))$, are not equidistant. The leading-order spacing, $\hbar\omega_{\rm cyc}[1-2\zeta(n+1)]$ with $\zeta=\hbar\omega_{\rm cyc}/E_g^*$, matches the $\xi_n$ of Ref.~\cite{Wu2026}; for GaAs at 1~T, $\zeta\approx9\times10^{-4}$. The cavity enters through $\hat{\boldsymbol\Pi}\to\hat{\boldsymbol\Pi}+e\mathbf A_c$ inside $F$, i.e., by Peierls substitution on the full ladder~\cite{Kim2026}. In the Landau basis, the derivative of $F(u(s))$ with respect to $s=eA_c/\sqrt{m^*\hbar\omega_{\rm cyc}}$ is a divided difference, $(\partial_sH)_{mn}=F[u_m,u_n](P_x)_{mn}$. The paramagnetic matrix elements are therefore $(E_{n+1}-E_n)/\hbar\omega_{\rm cyc}$ times their parabolic values (Ref.~\cite{Wu2026} uses the square root of this ratio). At integer filling $\nu$ only the $\nu-1\to\nu$ transition contributes.

The $\alpha$ independence of the relaxed energy, i.e., gauge invariance, then gives the \emph{generalized TRK sum rule}
\begin{equation}
D_{\rm eff}=D_{\rm par}\frac{\omega_\nu}{\omega_{\rm cyc}},\quad \Omega^2=D_{\rm eff}\,\omega_\nu,\quad \hbar\omega_\nu=E_\nu-E_{\nu-1}.
\label{eq:genTRK}
\end{equation}
For $B\to0$, $D_{\rm eff}\to D_{\rm par}F'(E_F)=D_{\rm par}v_F/v_F^{\rm par}$ to all orders, whose leading term is $D_{\rm par}[1+4\alpha_{\rm np}\langle E\rangle]$ with $\alpha_{\rm np}=-1/E_g^*$, as in Refs.~\cite{Kim2026,SqueezingDraft}. We verified Eq.~\eqref{eq:genTRK} numerically to $10^{-6}$ [Fig.~\ref{fig:kane}(c)]. Like the usual TRK sum rule, of which it is the nonparabolic generalization, Eq.~\eqref{eq:genTRK} is a consequence of gauge invariance and \emph{not} an independent assumption. It fixes the diamagnetic coefficient that a nonparabolic band must carry, and any model with nonparabolic level spacings but a parabolic $D$ violates it.

We evaluate the mean-field ground-state energy, treating the cavity field as a classical amplitude $\alpha$ (taken real) and filling the lowest $\nu$ levels of the single-particle Landau ladder,
\begin{equation}
\frac{E_G(\alpha)}{N}=\frac{\hbar\omega_{\rm cav}\alpha^2}{N}+\frac1\nu\sum_{n<\nu}\big[E_n(s)-E_n(0)\big],
\label{eq:EGmf}
\end{equation}
where $s=eA_c/\sqrt{m^*\hbar\omega_{\rm cyc}}$ is the cavity field in units of the Landau-level momentum scale. With $A_c=2A_0\alpha$ and $\hbar D=Ne^2A_0^2/2m^*$, it is related to the photon amplitude by $s^2=(8D/\omega_{\rm cyc})\,\alpha^2/N$. The $E_n(s)$ are the eigenvalues, in units of $\hbar\omega_{\rm cyc}$, of the ladder Hamiltonian $h(s)$ in one guiding-center sector; a uniform field couples only $n\to n\pm1$ within a sector, and all sectors are identical. We compare three choices of $h(s)$: the gauge-consistent one, $h(s)=F([(\hat\Pi_x+s)^2+\hat\Pi_y^2]/2)$; the model of Ref.~\cite{Wu2026}, with nonparabolic energies $\xi_n$ and matrix elements $f_n$, a parabolic diamagnetic term $s^2/2$, and the coupling scaled by $r=\Omega_{\rm eff}/\Omega_0$; and a ``project, then couple'' model, $h(s)=\mathrm{diag}\,F(n+\tfrac12)+s\hat\Pi_x+s^2/2$. Figure~\ref{fig:kane}(a,b) shows $E_G$ for $\nu=2$ and $D=0.25\,\omega_0$, with ladders truncated at 300--700 levels.

For the gauge-consistent ladder, $h(s)$ is unitarily equivalent to $h(0)$, because the shift $\hat\Pi_x\to\hat\Pi_x+s$ is a pure gauge. Its eigenvalues are therefore independent of $s$, and $E_G=\hbar\omega_{\rm cav}\alpha^2$, with the electronic part constant to within $10^{-14}$ (black curves). The model of Ref.~\cite{Wu2026} is not $\alpha$ independent. At the bare coupling ($r=1$) its energy still increases monotonically with $\alpha$, so the normal state remains the ground state (blue dashed). With the enhanced averaged coupling $\Omega_{\rm eff}=1.6\,\Omega_0$ ($r=1.6$), beyond its critical value, it develops minima at $\alpha/\sqrt N\approx1.5$ for $\zeta=0.02$ and $4.9$ for $\zeta=0.002$ (purple). The ``project, then couple'' model violates TRK already at linear order and can be unbounded below (dotted). Neither instability is present in the gauge-consistent model. In the model of Ref.~\cite{Wu2026}, nonparabolicity merely bounds an instability created by the enhanced averaged coupling, whereas in the ``project, then couple'' model the inconsistent coupling itself produces one.

Nonparabolicity therefore makes the problem weakly nonlinear, turning the dressed mode into a Kerr oscillator~\cite{SqueezingDraft}, but it cannot produce a condensate. The relaxed energy is independent of $\alpha$ for any band once the coupling is introduced by minimal substitution, so finite-field minima require a coupling that violates the (generalized) TRK sum rule, such as the enhanced averaged $\Omega_{\rm eff}$. This conclusion also holds classically, since a static uniform vector potential acting on a nonlinear charged fluid can be removed by a shift of the canonical momentum. What the modulation does produce is dynamical, namely the parametric creation of photon pairs from the vacuum, analyzed in Sec.~\ref{sec:finite}.

We note that the stability to all orders of Sec.~\ref{sec:positivity} is proven for the parabolic band, where the driven problem is exactly linear. With nonparabolicity, the quartic terms of the dispersion couple the c.m.\ and relative motions, with a relative strength of order $E_F/E_g^*\sim\nu\zeta$; this is small for GaAs ($\zeta\approx9\times10^{-4}$ at 1~T) and is reflected in the small deviation of $D_{\rm eff}/D_{\rm par}$ from unity in Fig.~\ref{fig:kane}(c). The driven dynamics is then no longer exactly linear, and Theorem~\ref{thm:krein} does not apply directly; the exclusion of a condensate, however, follows exactly from the gauge theorem for any band.

\section{High-frequency limit and the origin of the predicted transition}
\label{sec:highfreq}

\begin{figure*}[t]
\centering\includegraphics[width=\textwidth]{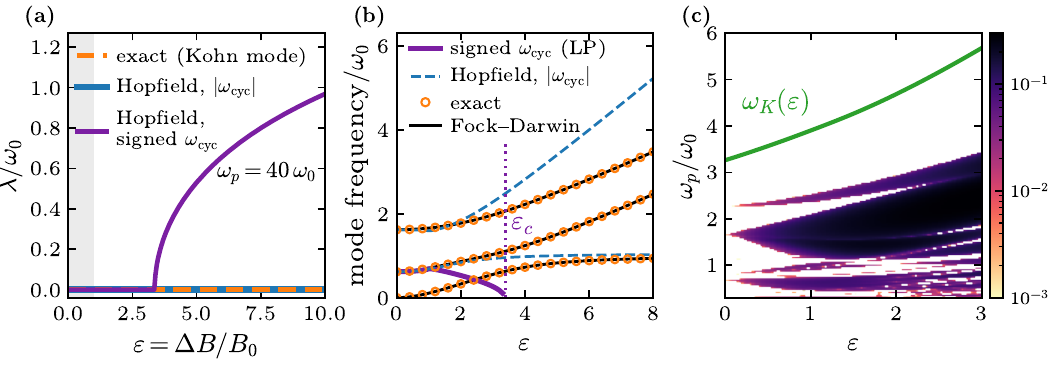}
\caption{Exact Floquet stability. (a)~Maximum Floquet exponent $\lambda$ at $\omega_p=40\,\omega_0$ (exact monodromy, no truncation). Purple: Hopfield model with signed $\omega_{\rm cyc}(t)$, unstable for $\varepsilon>3.4$. Blue: Hopfield model with $|\omega_{\rm cyc}(t)|$. Orange dashed: exact Kohn-mode model, Eq.~\eqref{eq:HK}. Gray: no field reversal. (b)~Mode frequencies at $\omega_p=40\,\omega_0$. Orange circles: exact quasienergies of Eq.~\eqref{eq:HK}. Black lines: Fock--Darwin model, Eq.~\eqref{eq:HKbar}. Blue dashed: time-averaged Hopfield model with $|\omega_{\rm cyc}|$. Purple: lower polariton (LP) of the signed model, which softens at $\varepsilon_c$. (c)~Lossless $(\varepsilon,\omega_p)$ map of $\lambda$ for Eq.~\eqref{eq:HK} with $\omega_d=0.2\,\omega_0$; green: Krein frequency $\omega_K(\varepsilon)$, Eq.~\eqref{eq:wK}, above which stability holds to all orders. Parameters: $\omega_{\rm cav}=\omega_0$, $\Omega_0=0.5\,\omega_0$ ($\eta=\Omega_0/\omega_{\rm cav}=0.5$, USC regime), $D=\Omega_0^2/\omega_0$; $\omega_d=0$ in (a,b).}
\label{fig:highfreq}
\end{figure*}

Averaging Eq.~\eqref{eq:HK} over a period, the terms linear in $B$ give the signed mean $\langle B\rangle=B_0$, while $(QA_B)^2$ gives $\langle B^2\rangle$. The result is
\begin{equation}
\begin{split}
\bar H_K={}&\frac{\big[\hat{\mathbf P}-Q\bar{\mathbf A}_B-QA_0(\aop+\ad)\mathbf e_x\big]^2}{2M}\\
&+\frac{M}{2}\Big(\omega_d^2+\frac{\sigma^2}{4}\Big)|\hat{\mathbf R}-\mathbf r_0|^2+\hbar\omega_{\rm cav}\ad\aop ,
\end{split}
\label{eq:HKbar}
\end{equation}
where $\bar{\mathbf A}_B$ corresponds to $B_0$ and $\sigma^2=\langle\omega_{\rm cyc}^2\rangle-\langle\omega_{\rm cyc}\rangle^2$. The added term is the ponderomotive (Kapitza) potential of the Faraday field~\cite{Kapitza1951}, $Q^2\langle|\mathbf E_F|^2\rangle/2M\omega_p^2$, with trap frequency $\omega_t=\varepsilon\omega_0/2\sqrt2$ for sinusoidal modulation. Equation~\eqref{eq:HKbar} describes a Fock--Darwin oscillator~\cite{Fock1928,Darwin1931} coupled to the cavity. Like Eq.~\eqref{eq:HK}, it is a sum of squares, and its coupling to the cavity is fixed by the charge and mass of the electrons, with $D$ unchanged. It therefore obeys the TRK sum rule exactly, so \emph{no polariton softens, and no SRPT occurs}.

Figure~\ref{fig:highfreq} confirms this at $\omega_p=40\,\omega_0$, the regime of Ref.~\cite{Wu2026}. In this regime the parameter-modulated model with signed $\omega_{\rm cyc}$ predicts an instability [purple curve in Fig.~\ref{fig:highfreq}(a)] and a lower polariton that softens to zero [purple curve in Fig.~\ref{fig:highfreq}(b)]. Neither occurs in the exact model, which includes the Faraday field (orange). Both positive models are stable for all $\varepsilon\le10$ [Fig.~\ref{fig:highfreq}(a)]; Eq.~\eqref{eq:wK} guarantees this up to $\varepsilon\approx30$. The signed model is unstable for $\varepsilon>3.4$, reproducing the threshold $c^2\varepsilon_c=1+\omega_0^2/4\Omega_0^2$ of Ref.~\cite{Wu2026}, with $c\simeq0.762$ defined by $\langle\sqrt{|1+\varepsilon\cos\omega_pt|}\rangle\simeq c\sqrt\varepsilon$ (called $\eta$ in Ref.~\cite{Wu2026}). The normal modes of Eq.~\eqref{eq:HKbar} match the exact quasienergies to within $2\times10^{-3}\omega_0$ for $\varepsilon\le8$ [Fig.~\ref{fig:highfreq}(b)]. A second-order van Vleck expansion converges to the exact quasienergies as $\omega_p^{-4}$ (Appendix~\ref{app:numerics}). Figure~\ref{fig:highfreq}(c) shows that every unstable point lies below $\omega_K$; the minimum gap is about $1\,\omega_0$.

This clarifies where the time-dependent Hopfield model works. It is a controlled approximation when the Landau basis follows the field adiabatically, $\omega_p\ll\omega_0$ with no field reversal ($\varepsilon<1$). There, the signed and $|\omega_{\rm cyc}|$ versions coincide, and both are stable. For $\omega_p\gg\omega_0$, Ref.~\cite{Wu2026} is right that the electrons respond to the signed mean field $\langle B\rangle$. However, the cavity couples to the kinetic momentum $\hat{\boldsymbol\Pi}$ with a strength that does not depend on $B$. The factor $\sqrt{|B|}$ in $\Omega$ is a Landau-basis matrix element, and its cycle average carries no physical enhancement. Averaging that matrix element separately from the Hamiltonian makes $\mathcal H(t)$ indefinite and opens the instability. The correct average, Eq.~\eqref{eq:HKbar}, replaces the enhanced coupling by a trap. The signed model's threshold, $\varepsilon_c\approx3.4$ for $\Omega_0=0.5\,\omega_0$, also lies in the field-reversal regime, where Landau quantization is lost during each cycle and where the Faraday field is very large (Sec.~\ref{sec:discussion}). The finite-field minima obtained in Ref.~\cite{Wu2026} with band nonparabolicity are excluded by the gauge theorem of Sec.~\ref{sec:gauge}.

\emph{The predicted SRPT as a gauge-invariance violation.} The mechanism can be stated in the language of the equilibrium no-go debate. Averaging the Hopfield parameters separately gives an effective static model with $\omega_{\rm eff}=\langle\omega_b\rangle$ and $\Omega_{\rm eff}=\Omega_0\langle\sqrt{|1+\varepsilon\cos\omega_pt|}\rangle$, with $D$ unchanged. For the signed model, $\omega_{\rm eff}=\omega_0$ is fixed while $\Omega_{\rm eff}$ grows as $\sqrt\varepsilon$, so the effective model violates the TRK sum rule, $\Omega_{\rm eff}^2>D\omega_{\rm eff}$, for $\varepsilon>1.54$. Its SRPT appears once the violation exceeds the photon stiffness, $\Omega_{\rm eff}^2-D\omega_{\rm eff}>\omega_{\rm cav}\omega_{\rm eff}/4$, which reproduces $\varepsilon_c=3.36$. This is the same violation that produced the spurious SRPT in models that drop or underweight the $A^2$ term. For the version with $|\omega_{\rm cyc}|$, by contrast, $\omega_{\rm eff}=\omega_0\langle|1+\varepsilon\cos\omega_pt|\rangle$, and Jensen's inequality\footnote{For a convex function $f$, Jensen's inequality states that $\langle f(x)\rangle\ge f(\langle x\rangle)$ for any average. With $f(x)=x^2$ and $x=\sqrt{|1+\varepsilon\cos\omega_pt|}$, it gives $\langle\sqrt{|1+\varepsilon\cos\omega_pt|}\rangle^2\le\langle|1+\varepsilon\cos\omega_pt|\rangle$, i.e., the square of an average never exceeds the average of the square (equivalently, the Cauchy--Schwarz inequality).} gives $\Omega_{\rm eff}^2=D\omega_0\langle\sqrt{|1+\varepsilon\cos\omega_pt|}\rangle^2\le D\omega_{\rm eff}$. TRK then holds as an inequality, and there is no transition.

Under driving there is a second, dynamical aspect of gauge invariance. A model can satisfy TRK at every instant and still fail to be gauge invariant. A time-dependent $B(t)$ is described by $\mathbf A_B(\mathbf r,t)$, and the gauge-invariant electric field $\mathbf E=-\partial_t\mathbf A_B-\nabla\phi$ is the Faraday field, which Faraday's law requires. In the instantaneous Landau basis, the same physics appears as the term $-i\hbar\hat U_t^\dagger\dot{\hat U}_t\propto\dot B$ generated by the time-dependent basis change, i.e., by a time-dependent unitary (gauge) transformation (Sec.~\ref{sec:staticbasis}). The parameter-modulated model omits both, which is equivalent to setting $\mathbf E=0$, inconsistent with Faraday's law. Instantaneous TRK is therefore necessary but not sufficient for a gauge-invariant driven model. A related conclusion was reached recently for Floquet physics with quantized light, where gauge-consistent descriptions reproduce the Floquet quasienergies throughout parameter space while inconsistent truncations do not~\cite{PerezGonzalez2026}. This dynamical violation is why both parameter-modulated versions miss the Faraday tongues and the Fock--Darwin high-frequency limit, even the version that is stable.

\section{Finite drive frequency: tongues, loss, and dynamical Casimir emission}
\label{sec:finite}

\begin{figure*}[t]
\centering\includegraphics[width=\textwidth]{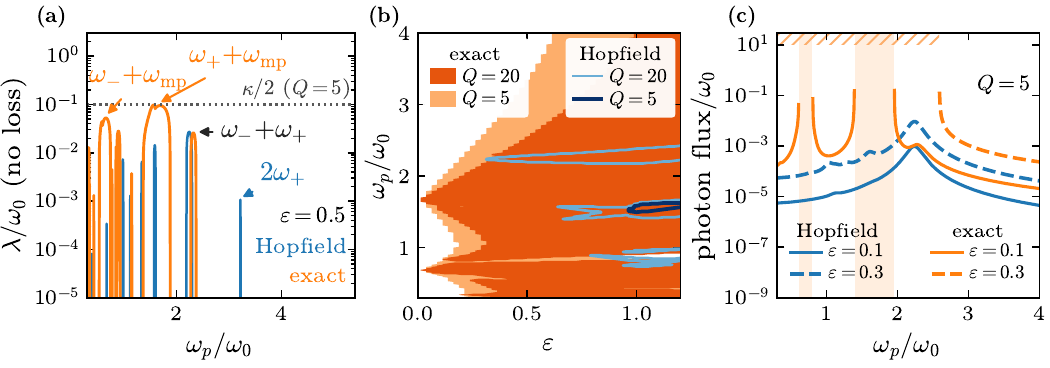}
\caption{Tongues, thresholds, and emission in the USC regime ($\eta=0.5$). (a)~Largest Floquet exponent without loss versus $\omega_p$ at $\varepsilon=0.5$, for the Hopfield model with $|\omega_{\rm cyc}|$ (blue) and the exact Kohn model with $\omega_d=0.2\,\omega_0$ (orange); $\Delta\omega_p=10^{-3}\omega_0$. Each peak is an instability tongue crossed at this depth; arrows mark the labeled ones: the polariton pair $\omega_-+\omega_+$ (both models), $2\omega_+$ (Hopfield), and the Faraday-driven pairs of a polariton and the low magnetoplasmon, $\omega_\pm+\omega_{\rm mp}$ (exact model only). Unlabeled peaks are dressed and higher-order ($n\ge2$) resonances. Dotted: $\kappa/2$ for $Q=5$, a guide to which tongues survive loss. (b)~Regions where the system with loss is unstable, for cavity quality factor $Q=\omega_0/\kappa_0$ and Drude damping $\gamma=0.01\,\omega_0$. Orange regions: exact Kohn model (dark: $Q=20$; light: $Q=5$); the $Q=20$ region lies almost entirely inside the $Q=5$ region. Blue contours: Hopfield model (light: $Q=20$; dark: $Q=5$). (c)~Emitted photon flux, Eq.~\eqref{eq:flux}, for $Q=5$. Solid: $\varepsilon=0.1$; dashed: $\varepsilon=0.3$. The flux diverges on approaching threshold. Shaded bands ($\varepsilon=0.1$) and the hatched bar ($\varepsilon=0.3$) mark where the exact model is above threshold and the linear model has no steady state.}
\label{fig:tongues}
\end{figure*}

\subsection{Lossless tongue structure}
Reference~\cite{Wu2026} anticipated that driving within the polariton spectrum leads to parametric amplification and polariton pair creation.
Here we analyze this regime in full, including the Faraday field. 

Figure~\ref{fig:tongues}(a) resolves the tongues at $\varepsilon=0.5$, in the USC regime with $\eta=0.5$, as throughout. The Hopfield model with $|\omega_{\rm cyc}|$ has a hierarchy of narrow sum resonances. 
The strongest is at $\omega_-+\omega_+=2.243\,\omega_0$, with $\lambda=0.027\,\omega_0$, in close agreement with the first-order estimate $|K^{(1)}_{-+}|=0.028\,\omega_0$ of Eq.~\eqref{eq:rates}. Weaker tongues follow at $\omega_p\approx1.59\,\omega_0$ (the second-order resonance $2\omega_+=2\omega_p$), at $1.34\,\omega_0$ (the dressed $2\omega_-$), and at $3.22\,\omega_0$ ($2\omega_+$), together with higher-order and combination resonances.
The exact model adds broad tongues at $\omega_p\approx\omega_\pm+\omega_{\rm mp}$, where $\lambda$ reaches $\approx0.1\,\omega_0$. These tongues are generated by the Faraday field, which couples the c.m.\ position to its kinetic momentum [the guiding-center terms of Eq.~\eqref{eq:static_basis}]. Without the cavity they reduce to a Mathieu instability of the Larmor-frame motion, $\ddot{\mathbf R}=-\tfrac14\omega_{\rm cyc}(t)^2\mathbf R$, at $\omega_p\simeq\omega_0$ with $\lambda\simeq\varepsilon\omega_0/4$. With the cavity this splits into the polariton--magnetoplasmon tongues. The $2\omega_\pm$ tongues vanish at first order in the exact model, while $\omega_-+\omega_+$ survives with the Hopfield strength.

\emph{Role of ultrastrong coupling.} The two families of tongues depend on the coupling in different ways. Creating a pair of polaritons \emph{requires the counter-rotating terms of the light--matter interaction}, which mix creation and annihilation operators in the dressed modes. The first-order pair coupling of the $\omega_-+\omega_+$ tongue therefore scales as $\eta^2$. We find $|K^{(1)}_{-+}|/\varepsilon=0.0006$, $0.0025$, $0.0098$, and $0.056\,\omega_0$ for $\eta=0.05$, 0.1, 0.2, and 0.5, so this channel, the source of two-mode squeezed photon pairs, is a genuine USC effect, which is negligible at weak coupling. The Faraday tongues are different. Their pair coupling, $\approx0.19$--$0.20\,\varepsilon\,\omega_0$, hardly depends on $\eta$, because they originate in the driven c.m.\ motion itself, a matter parametric resonance that exists even without the cavity. The cavity determines how much of this pair creation is radiated as photons, through the photonic content of the polariton partner. In the USC regime, the polaritons are strongly mixed and widely split ($\omega_+-\omega_-\simeq2\Omega_0$), so both families of tongues are well separated and efficiently radiate. More fundamentally, the question of a Floquet SRPT is itself a USC question, since an SRPT could only occur at couplings comparable to the bare frequencies, which is where the no-go theorem, and its dynamical extension here, matter.

\subsection{Open-system theory: cavity loss and Drude damping}
\label{sec:open}

In the USC regime, dissipation must be formulated in terms of dressed operators~\cite{BeaudoinGambettaBlais2011}. We use a dressed master equation on the static (undriven) normal modes $\hat c_k=\ell_k^{\rm T}\hat X$, obtained by symplectic (Bogoliubov) diagonalization of $\mathcal H$ at $\varepsilon=0$. The index $k$ labels these modes as the lower and upper polaritons ($k=\mp$) for the Hopfield model, and, for the exact model with $\omega_d>0$, the two polaritons together with the low magnetoplasmon ($k=-,+,{\rm mp}$), corresponding to the three degrees of freedom: the cavity mode and the two c.m.\ coordinates $X$ and $Y$. Each mode decays through its own Lindblad operator $\sqrt{\Gamma_k}\,\hat c_k$, with
\begin{equation}
\Gamma_k=\kappa_0|u_k|^2+2\gamma f_k^{\rm kin},\qquad \aop=\sum_k(u_k\hat c_k+v_k\hat c_k^\dagger),
\label{eq:rates_diss}
\end{equation}
where $u_k$ and $v_k$ are the Bogoliubov coefficients of the photon operator, so $|u_k|^2$ is the photonic weight of mode $k$.

The two terms in $\Gamma_k$ have different origins. The first is cavity loss. It follows from the quantized quasinormal-mode theory, in which the cavity couples to its reservoir through $\hat\Pi_c=\aop$~\cite{GustinPRL2025,LSA2026}, so that each dressed mode leaks at a rate proportional to its photonic weight; at $B=0$ this gives the decay rate $\kappa_0\cosh^2r$ of the dressed mode used in Ref.~\cite{SqueezingDraft}. For the Hopfield polaritons, $|u_-|^2=0.29$ and $|u_+|^2=0.77$. The second term is Drude damping of the electrons, modeled as a friction force $-M\gamma\dot{\mathbf R}$ on the c.m.\ that dissipates kinetic energy at the rate $2\gamma\,\hat{\boldsymbol\Pi}^2/2M$.

For weak damping, the energy of a normal mode then decays at $2\gamma$ times the fraction of its energy that is electronic kinetic energy, $f_k^{\rm kin}=\langle\hat{\boldsymbol\Pi}^2/2M\rangle_k/\langle\hat H\rangle_k$, evaluated in the undamped mode; this is the standard first-order (dissipative-function) result for weakly damped coupled oscillators~\cite{LandauMechanics}. A bare cyclotron mode, whose energy is entirely kinetic, therefore has linewidth $2\gamma$, independent of $B$. By contrast, the low magnetoplasmon stores its energy mostly as potential energy of the restoring force, so its Drude linewidth is strongly reduced, which is the origin of the small $\Gamma_{\rm mp}$ discussed below.

Since the Hamiltonian is quadratic and the Lindblad operators are linear in $\hat X$, the master equation maps Gaussian states onto Gaussian states. Starting from the vacuum, all first moments vanish, and the state is fully characterized by the covariance matrix $\Sigma_{ij}=\tfrac12\langle\{\Delta\hat X_i,\Delta\hat X_j\}\rangle$, which contains all photon numbers, correlations, and squeezing. Writing the Lindblad operators as $\sqrt{\Gamma_k}\,\hat c_k=\sqrt{\Gamma_k}\,\ell_k^{\rm T}\hat X$ and collecting them in the matrix $C=\sum_k\Gamma_k\ell_k^*\ell_k^{\rm T}$, the master equation gives a closed equation for $\Sigma$,
\begin{equation}
\begin{split}
\dot\Sigma&=A\Sigma+\Sigma A^{\rm T}+\mathcal D,\\
A&=J[\mathcal H(t)+{\rm Im}\,C],\quad \mathcal D=J\,{\rm Re}\,C\,J^{\rm T}.
\end{split}
\label{eq:lyap}
\end{equation}
The matrix $A$ governs the deterministic evolution of the mean quadratures, $\langle\dot{\hat X}\rangle=A\langle\hat X\rangle$, and contains both the driven Hamiltonian dynamics, $J\mathcal H(t)$, and the damping, $J\,{\rm Im}\,C$; it is called the drift matrix in the language of Fokker--Planck equations. The matrix $\mathcal D$ is the corresponding diffusion (noise), which is required by the fluctuation--dissipation relation and keeps the steady state at or above the vacuum level.

The steady state at $\varepsilon=0$ is the dressed vacuum, so the ground-state virtual photons are not emitted. Under driving, integrating Eq.~\eqref{eq:lyap} over one period gives the map $\Sigma\mapsto\Phi_A\Sigma\Phi_A^{\rm T}+\mathcal Q$, where $\Phi_A$ is the one-period propagator of $A$ and $\mathcal Q=\int_0^T\Phi_A(T,t)\,\mathcal D\,\Phi_A(T,t)^{\rm T}dt$ is the noise accumulated over one period. Below threshold, all eigenvalues of $\Phi_A$ lie inside the unit circle, and the periodic (Floquet) steady state is the fixed point $\Sigma_{\rm ss}=\sum_{n\ge0}\Phi_A^n\mathcal Q\,(\Phi_A^{\rm T})^n$, which we sum by repeated doubling. The emitted photon flux is
\begin{equation}
\Phi_{\rm ph}=\kappa_0\sum_k|u_k|^2\,\overline{\langle\hat c_k^\dagger\hat c_k\rangle},
\label{eq:flux}
\end{equation}
where the overline denotes the period average. This dissipator treats the Hamiltonian exactly, without a rotating-wave approximation, but it is secular and built on the static modes rather than Floquet resolved~\cite{AkbariFloquet2026,AkbariFLT2026}. We tested both simplifications (Appendix~\ref{app:open}). A non-secular cavity channel leaves the thresholds unchanged. Fluxes at resonances change by less than 10\% for the Hopfield model and by up to a factor of 3 near threshold in the exact model. Only the off-resonant background is sensitive to the choice of modes. The resonant features and thresholds, and hence all qualitative predictions, are robust.

\begin{figure*}[t]
\centering\includegraphics[width=\textwidth]{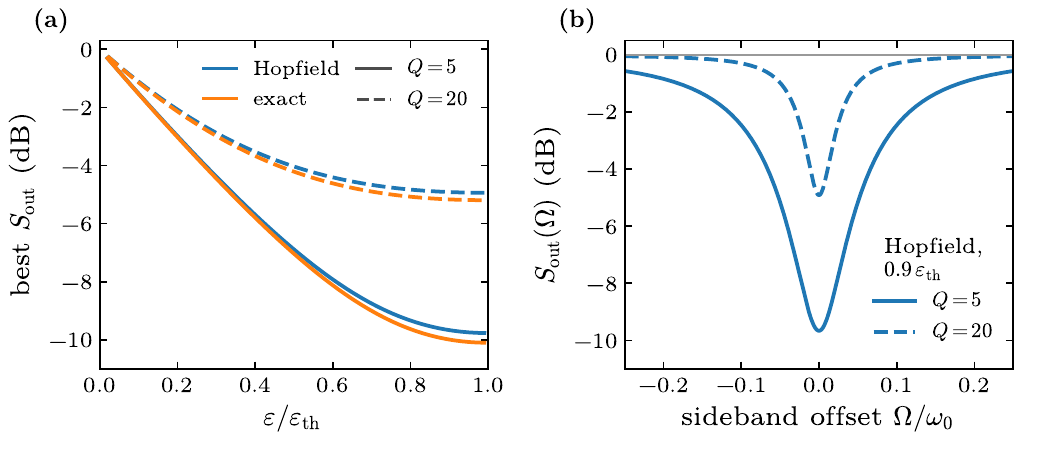}
\caption{Two-mode output squeezing at the $\omega_-+\omega_+$ tongue in the USC regime ($\eta=0.5$). (a)~Best $S_{\rm out}$ (minimized over quadrature and sideband offset) versus $\varepsilon/\varepsilon_{\rm th}$ for the Hopfield (blue) and exact (orange) models, $Q=5$ (solid) and 20 (dashed); $\gamma=0.01\,\omega_0$. (b)~Spectrum versus sideband offset $\Omega$ at $0.9\,\varepsilon_{\rm th}$ (Hopfield).}
\label{fig:squeezing}
\end{figure*}

\subsection{Instability thresholds with loss, anti-damping, and sample dependence}
For a nondegenerate pair with rates $\Gamma_i$ and $\Gamma_j$ and detuning $\Delta$ from resonance, the rotating-wave model gives instability when
\begin{equation}
|K^{(1)}_{ij}|^2>\tfrac14\Gamma_i\Gamma_j\big[1+4\Delta^2/(\Gamma_i+\Gamma_j)^2\big].
\label{eq:threshold}
\end{equation}
For the Hopfield tongues this requires $\varepsilon\approx1$ ($Q=5$) or $\varepsilon\approx0.33$ ($Q=20$) [Fig.~\ref{fig:tongues}(b)]. The Faraday tongues pair a polariton with the magnetoplasmon, which with Drude damping alone is almost undamped ($\Gamma_{\rm mp}\approx10^{-3}\omega_0$), because its energy is mostly potential rather than kinetic. Three effects follow. First, the threshold $\propto\sqrt{\Gamma_\pm\Gamma_{\rm mp}}$ is low, $\varepsilon\approx0.05$ for $Q=5$ near $\omega_-+\omega_{\rm mp}$. Second, for $\Gamma_j\ll\Gamma_i$ the detuning penalty in Eq.~\eqref{eq:threshold} is $\simeq\Gamma_j\Delta^2/\Gamma_i$, which decreases as the polariton linewidth grows. Cavity loss can therefore destabilize detuned points that are stable without loss, because the lossy polariton mediates anti-damping of the magnetoplasmon, as for blue-detuned driving in cavity optomechanics~\cite{Aspelmeyer2014}. A lossier cavity accordingly enlarges the unstable region, which explains why the $Q=20$ region lies almost entirely inside the $Q=5$ one in Fig.~\ref{fig:tongues}(b). Third, because $\omega_{\rm mp}$ is small, tongues of different order lie close together. Together these effects merge the Faraday tongues into a near continuum.

This is strongly sample dependent. Edge magnetoplasmons of real samples have additional radiative, edge-scattering, and inhomogeneous damping. Table~\ref{tab:gmp} shows how the unstable area for $\varepsilon\le0.6$ and $Q=5$, and the minimum threshold, change when we add damping to the magnetoplasmon by hand.

\begin{table}[t]
\caption{Sensitivity of the Faraday-tongue instability to the magnetoplasmon damping $\Gamma_{\rm mp}$ ($Q=5$, $\gamma=0.01\,\omega_0$, $\varepsilon\le0.6$).}
\label{tab:gmp}
\begin{ruledtabular}
\begin{tabular}{lcc}
$\Gamma_{\rm mp}/\omega_0$ & unstable area & minimum $\varepsilon_{\rm th}$\\
\hline
0.0008 (Drude only) & 61\% & 0.05\\
0.011 & 20\% & 0.13\\
0.031 & 11\% & 0.20\\
0.101 & 6\% & 0.33\\
\end{tabular}
\end{ruledtabular}
\end{table}
With realistic damping the continuum breaks up into discrete Faraday tongues with thresholds $\varepsilon\approx0.1$--$0.3$. Between neighboring tongues, genuinely stable windows remain, e.g., at $\omega_p\approx0.85$--$0.91\,\omega_0$ for $\varepsilon=1$ and $Q=5$.

To be clear, 
\emph{none of these instabilities is an SRPT}. Above threshold, the modulation pumps energy into the c.m.\ motion of the electron gas, a betatron-like magnetic pumping of the guiding center, and into the polaritons. This is driven-dissipative parametric growth, not a condensate in a Floquet ground state.

\subsection{Photon emission and two-mode squeezing}
Below threshold the system emits DCE photons [Fig.~\ref{fig:tongues}(c)]. At $\omega_0/2\pi=0.5$~THz, a flux of $10^{-3}\omega_0$ corresponds to about $3\times10^9$ photons/s. Near threshold the flux grows as $1/|\lambda|$, where $|\lambda|$ is the decay rate of the least-damped Floquet mode, which vanishes at threshold, as for a laser or an optical parametric oscillator just below threshold.

Whether this light is squeezed depends on the tongue. In a Faraday tongue, the partner of each photon is a magnetoplasmon, which does not radiate. The emitted light is then correlated with the electronic motion but, on its own, thermal, since a single mode of a two-mode squeezed state is thermal. The $\omega_-+\omega_+$ tongue instead creates pairs in which both partners are partly photonic. Its output consists of two-color photon pairs at $\omega_++\Omega$ and $\omega_--\Omega$, symmetric about $\omega_p/2$, which are two-mode squeezed, as in a nondegenerate optical parametric amplifier.

\emph{What is measured.} Squeezing refers to quantum fluctuations of a field quadrature below the vacuum level. For the two-color output here, the relevant quantity is two-mode squeezing. The output fields near $\omega_-$ and $\omega_+$ carry correlated quantum noise, so that a suitable combination of their quadratures has noise below the vacuum level, even though each field on its own is noisier than vacuum. Experimentally, this is the noise spectrum of the combined quadrature measured by homodyne detection, with local oscillators phase locked to the modulation. It combines the spectral components of the output at $\omega_++\Omega$ and $\omega_--\Omega$, where $\Omega$ is the sideband offset. We characterize it by the normalized noise spectrum $S_{\rm out}(\Omega)=10\log_{10}N(\Omega)$, where $N=1$ (0~dB) is the vacuum level and $N<1$ (negative dB) indicates squeezing. The squeezing is largest at $\Omega\approx0$, i.e., at the polariton frequencies themselves, and its bandwidth is set by the polariton linewidths [Fig.~\ref{fig:squeezing}(b)].

\emph{How it is computed.} We need the output fields in the periodic steady state. All USC effects, including the counter-rotating terms, are contained in the static dressed polariton operators $\hat c_\mp$, obtained by exact (Bogoliubov) diagonalization of the Hopfield or Kohn Hamiltonian. The output fields follow from the dressed-state input--output relations, $\hat b^{\rm out}_\mp=\hat b^{\rm in}_\mp-\sqrt{\kappa^{\rm rad}_\mp}\,\hat c_\mp$~\cite{GustinPRL2025,LSA2026,SqueezingDraft}, with radiative rates $\kappa^{\rm rad}_\mp=\kappa_0|u_\mp|^2$. The polariton linewidths $\Gamma_\mp$ contain these radiative parts and the Drude parts $\Gamma_\mp-\kappa^{\rm rad}_\mp$, which are absorbed and not emitted. The modulation, by contrast, is weak compared with the polariton frequencies near threshold ($G\ll\omega_\pm,\,\omega_+-\omega_-$).

We therefore treat the modulation in the rotating-wave approximation (RWA) with respect to the dressed modes, keeping only the resonant pair term $G(\hat c_-^\dagger\hat c_+^\dagger e^{-i\omega_pt}+{\rm H.c.})$ and dropping terms that oscillate at $\omega_p$ and its harmonics. This RWA concerns the drive, not the light--matter coupling, which is treated exactly. The pair coupling is $G=|K^{(1)}_{-+}|\simeq0.056\,\varepsilon\,\omega_0$ (Hopfield; $0.053\,\varepsilon\,\omega_0$ exact), and $\Delta=\omega_-+\omega_+-\omega_p$ is the detuning. In the frame rotating at the polariton frequencies, the Fourier components $u(\Omega)=(\hat c_-(\Omega),\hat c_+^\dagger(\Omega))^{\rm T}$ obey $\mathsf M(\Omega)u=K_r^{1/2}b^{\rm in}_{\rm rad}+K_a^{1/2}b^{\rm in}_{\rm abs}$, with
\begin{equation}
\mathsf M(\Omega)=\begin{pmatrix}\Gamma_-/2+i(\Delta-\Omega)&iG\\-iG&\Gamma_+/2-i\Omega\end{pmatrix},
\end{equation}
where $K_{r,a}={\rm diag}(\kappa^{(r,a)}_-,\kappa^{(r,a)}_+)$ contain the radiative and absorptive rates, and $b^{\rm in}_{\rm rad,abs}$ are the corresponding vacuum input noises. 

Inserting this into the input--output relations gives the output in terms of the inputs, $(\hat b^{\rm out}_-(\Omega),\hat b^{\rm out\dagger}_+(\Omega))^{\rm T}=T^{\rm rad}b^{\rm in}_{\rm rad}+T^{\rm abs}b^{\rm in}_{\rm abs}$, with the $2\times2$ transfer matrices
\begin{equation}
T^{\rm rad}=\openone-K_r^{1/2}\mathsf M^{-1}K_r^{1/2},\qquad T^{\rm abs}=-K_r^{1/2}\mathsf M^{-1}K_a^{1/2}.
\end{equation}
The element $T^{(j)}_{mk}$ gives the contribution of the input noise of channel $j$ (radiative or absorptive) of mode $k$ ($k=1$: lower, $k=2$: upper polariton) to output component $m$ ($m=1$: $\hat b^{\rm out}_-$; $m=2$: $\hat b^{\rm out\dagger}_+$). The output noise of the weighted two-mode quadrature $\cos\phi\,e^{-i\theta}\hat b^{\rm out}_-+\sin\phi\,e^{i\theta}\hat b^{\rm out\dagger}_+$ is then
\begin{equation}
N_{\theta\phi}(\Omega)=\sum_{j={\rm rad,abs}}\sum_{k=1,2}\big|\cos\phi\,e^{-i\theta}T^{(j)}_{1k}+\sin\phi\,e^{i\theta}T^{(j)}_{2k}\big|^2 ,
\end{equation}
with $N=1$ for vacuum inputs and no modulation.

We quote $S_{\rm out}=10\log_{10}\min_{\theta,\phi}N$, minimized over the quadrature angle $\theta$ (the local-oscillator phase) and the mixing angle $\phi$, which sets the relative weight of the two output colors and compensates for the unequal linewidths of the two polaritons; the minimization over $\phi$ is a $2\times2$ eigenvalue problem. The measurement requires a local oscillator phase locked to the modulation, e.g., two-color homodyne detection or sub-cycle electro-optic sampling~\cite{Riek2015,BeneaChelmus2019}. The RWA treatment of the drive is accurate. It reproduces the intracavity two-mode squeezing of the exact non-RWA Floquet covariance, $-0.39$ ($-1.04$)~dB at $\varepsilon=0.1$ ($0.3$), compared with $-0.40$ ($-1.08$)~dB, and the thresholds agree with the full maps.

\begin{table}[th]
\caption{Two-mode output squeezing at the $\omega_-+\omega_+$ tongue.}
\label{tab:squeezing}
\begin{ruledtabular}
\begin{tabular}{lcccc}
model, $Q$ & $\varepsilon_{\rm th}$ & $\kappa^{\rm rad}_\mp/\Gamma_\mp$ & $S_{\rm out}(0.5\,\varepsilon_{\rm th})$ & $S_{\rm out}(0.9\,\varepsilon_{\rm th})$\\
\hline
Hopfield, 5 & 0.95 & 0.85, 0.94 & $-6.8$~dB & $-9.7$~dB\\
Hopfield, 20 & 0.31 & 0.59, 0.79 & $-4.0$~dB & $-4.9$~dB\\
exact, 5 & 1.00 & 0.87, 0.94 & $-7.0$~dB & $-10.0$~dB\\
exact, 20 & 0.32 & 0.63, 0.79 & $-4.2$~dB & $-5.2$~dB\\
\end{tabular}
\end{ruledtabular}
\end{table}

Figure~\ref{fig:squeezing} and Table~\ref{tab:squeezing} summarize the results. The squeezing reaches $-4$ to $-7$~dB at half threshold and $-5$ to $-10$~dB near threshold, over bandwidths of $0.04$--$0.1\,\omega_0$ (20--50~GHz at 0.5~THz). As for the Kerr squeezing of Ref.~\cite{SqueezingDraft}, a higher $Q$ lowers the threshold but reduces the radiative efficiency, because the Drude rate is fixed, and the squeezing is bounded by $\kappa^{\rm rad}/\Gamma$. Unlike the Kerr mechanism, the squeezing here starts linearly in $\varepsilon$ and needs no coherent intracavity field, since below threshold the intracavity occupation is of order one. The Hopfield and exact models give nearly identical squeezing. The $\omega_-+\omega_+$ tongue involves only the two polaritons in both models, with almost the same pair coupling (0.056 versus $0.053\,\varepsilon\,\omega_0$) and radiative efficiencies. Since $S_{\rm out}$, as a function of $\varepsilon/\varepsilon_{\rm th}$, depends essentially only on these efficiencies, the two curves nearly coincide. The magnetoplasmon, which distinguishes the exact model, does not take part in this channel; its effect is to add the competing Faraday instabilities discussed above.

These numbers come from an idealized model, and three caveats apply. (i)~The thresholds require $\varepsilon\approx0.3$--$1$ at $\omega_p\approx2.3\,\omega_0$, which implies Faraday fields of $3\times10^6$--$10^7$~V/m (Sec.~\ref{sec:discussion}). Heating will reduce the squeezing. (ii)~In the exact model with Drude-only magnetoplasmon damping, the $\omega_-+\omega_+$ tongue at $Q=5$ is preempted by Faraday-type instabilities from $\varepsilon\approx0.2$; with $\Gamma_{\rm mp}\gtrsim0.03\,\omega_0$, or at $Q=20$, it is not. (iii)~Thermal noise in the absorptive channel is neglected.

\subsection{Above threshold}
Above threshold the linear model has no steady state. In reality the growth is clamped, as in a laser or an optical parametric oscillator. Kane nonparabolicity shifts the modes out of the tongue, giving phase-locked parametric oscillation, but only at $n_c\sim\lambda/|\chi_{\rm K}|\sim10^6$--$10^7$ photons, with $|\chi_{\rm K}|/\omega_{\rm cav}\approx5\times10^{-9}$~\cite{SqueezingDraft}. This corresponds to quiver energies beyond the LO-phonon runaway ($U_p\approx10$--15~meV, $n_c\sim10^6$)~\cite{SqueezingDraft,Kim2026}, so heating, not coherent nonlinearity, limits the amplitude. Loading of the modulating circuit also contributes. Finally, for a modulation of duration $t_p$ the photon number grows by at most $e^{2\lambda t_p}$. With $\lambda=0.1\,\omega_0$ at $\omega_0/2\pi=0.5$~THz and $t_p=1$~ps, as proposed in Ref.~\cite{Wu2026}, $2\lambda t_p\approx0.6$, i.e., less than a factor of 2.

\section{Generality: modulated linear ultrastrong coupling}
\label{sec:generality}

Three features are specific to modulating $B$: the Faraday field, the guiding-center mode with its Fock--Darwin high-frequency limit, and field reversal. Apart from these, the positivity and Krein arguments use only two properties: the Hamiltonian is quadratic, and it is obtained by minimal coupling. They therefore apply to any linear, minimally coupled USC system. Examples include Lorentz-oscillator media, intersubband transitions, and phonon polaritons, all of which are Hopfield models with an exact classical counterpart~\cite{Hopfield1958,Ciuti2005,ChangPRA2025,Muniain2025}.

The averaging pitfall is equally general. Consider $N$ charges $q$ of mass $m$ on springs with a modulated stiffness, coupled to a uniform cavity field. In c.m.\ variables, with $Q=Nq$ and $M=Nm$,
\begin{equation}
\hat H(t)=\frac{[\hat P-QA_0(\aop+\ad)]^2}{2M}+\frac{M\omega_b(t)^2}{2}\hat X^2+\hbar\omega_{\rm cav}\ad\aop .
\label{eq:lorentz}
\end{equation}
The coupling, $-(QA_0/M)(\aop+\ad)\hat P$, and the diamagnetic term, $(Q^2A_0^2/2M)(\aop+\ad)^2$, do not depend on $\omega_b$. In the static basis, $\hat P=i\sqrt{\hbar M\omega_b/2}\,(\bd-\bop)$, which gives the Hopfield model with $\Omega^2=D\omega_b$ and $\hbar D=Q^2A_0^2/2M$. The factor $\sqrt{\omega_b}$ comes from the basis. 

Since Eq.~\eqref{eq:lorentz} is a sum of squares at every instant, as long as $\omega_b^2>0$, positivity and Theorem~\ref{thm:krein} apply. The canonical high-frequency average gives the same Hamiltonian with $\omega_b^2\to\langle\omega_b^2\rangle$ and unchanged coupling, i.e., a Hopfield model with $\bar\omega_b=\sqrt{\langle\omega_b^2\rangle}$ and $\bar\Omega^2=D\bar\omega_b$, which obeys TRK and is stable. Averaging the parameters of the fixed-basis Hopfield model instead gives $\omega_b\to\langle\omega_b\rangle$ and $\Omega\to\sqrt D\langle\sqrt{\omega_b}\rangle$. That procedure differs from the canonical one and neglects the basis-rotation terms $\propto(\dot\omega_b/\omega_b)(\bop^2-\bd{}^2)$, the analog of the Faraday terms of the Landau problem. For $\omega_b(t)^2=\omega_0^2(1+\varepsilon\cos\omega_pt)$, for example, the canonical average gives $\bar\omega_b=\omega_0$, whereas the parameter average gives $\langle\omega_b\rangle=\omega_0\langle\sqrt{1+\varepsilon\cos\omega_pt}\rangle<\omega_0$. Modulation with $\omega_b^2<0$ during part of the cycle makes the oscillator itself unstable; an inverted potential is a matter instability, not an SRPT.

The general rule is therefore to modulate and average the minimal-coupling Hamiltonian, not the parameters of its projection onto a particular basis. Positivity, stability to all orders above $\omega_K$, and pair creation as the only instability below it then follow for every such system. These conclusions are consistent with full Maxwell--Lorentz simulations of modulated linear dipoles~\cite{PyCharge,ChangPRA2025}, and with time-varying photonics more broadly. A time-invariant linear medium has a causal, passive response $\chi(t-t')$, whereas a modulated medium has $\chi(t,t')$ and allows parametric gain fed by the drive~\cite{Galiffi2022,Lustig2018}. These statements are restricted to linear, minimally coupled matter. Two-level (quantum Rabi or Dicke) matter, such as the modulated Dicke model of Ref.~\cite{HuangTian2022}, and nonlinear corrections beyond lowest-order Holstein--Primakoff lie outside their scope, and whether driving can induce a transition there remains open. For gauge-invariantly truncated multilevel models~\cite{DiStefano2019,Garziano2020} in which the modulation acts on the matter Hamiltonian alone, the unitary argument of Sec.~\ref{sec:gauge} still excludes a static mean-field condensate; a full analysis including saturation and quantum fluctuations is left for future work.

Magnetic (Zeeman) coupling is the exception, and it is worth being precise about why. For spins coupled to the cavity magnetic field, the Pauli Hamiltonian contains $(\hat{\mathbf p}+e\hat{\mathbf A})^2$ for the orbital motion but only $\boldsymbol\sigma\cdot\hat{\mathbf B}$ for the spin, so there is no diamagnetic protection and the TRK-based no-go theorem does not apply in the same form. Photon condensation is then possible in principle~\cite{RomanRoche2021}, and Floquet-engineered transitions have been proposed in cavity magnonics~\cite{Lin2026}. This is not a breakdown of Maxwell's equations. A \emph{linear} (harmonic) magnetic dipole has a bounded static susceptibility and, coupled to Maxwell's equations, is as stable as an electric Lorentz oscillator. 

The ingredient beyond a linear model is the \emph{matter}. A spin ensemble has a Curie susceptibility $\chi\propto1/T$, which grows without bound at low temperature, and the spins saturate once ordered, so the ordered phase requires nonlinear (Holstein--Primakoff) terms. The transition is a magnetostatic ordering of nonlinear spins, i.e., dipolar ferromagnetism mediated by the quasi-static field of the cavity, not the condensation of a radiative photon mode. It parallels the electric case of Ref.~\cite{Lamberto2025} and the Condon instability of Ref.~\cite{Andolina2020}; in all these cases the instability belongs to the matter and is enabled by a large static susceptibility together with nonlinearity. \emph{For linear matter, electric or magnetic, Maxwell's equations forbid an equilibrium, radiation-induced condensate, with or without periodic modulation.}

\section{Discussion}
\label{sec:discussion}

\emph{Physical scales.} Table~\ref{tab:params} in Appendix~\ref{app:params} converts the dimensionless parameters to physical
 units for $\omega_0/2\pi=0.5$~THz ($B_0\approx1.2$~T in GaAs). The coupling, $D/\omega_0=0.25$, is close to that of the sample of Refs.~\cite{Kim2026,SqueezingDraft} ($D/\omega_{\rm cav}=0.22$). The Faraday field at distance $r$ from the flux axis is $E\sim r\omega_p\Delta B/2$. At the operating point of Ref.~\cite{Wu2026} ($\Delta B=5$~T, $\omega_p/2\pi=20$~THz, $r=2.5~\mu$m), $E\sim10^9$~V/m. At the Faraday-tongue thresholds, $E\sim10^5$--$10^6$~V/m, and at the $\omega_-+\omega_+$ thresholds, $E\sim3\times10^6$--$10^7$~V/m. All are comparable to or above the kV/cm fields that already heat GaAs electron gases~\cite{Kim2026}. In the parabolic model the Faraday field heats only the relative motion, which decouples from the cavity; nonparabolicity and disorder couple it back. Below threshold the photon numbers are of order one per mode, so heating is dominated by the modulation, not by the emitted light.

\emph{Refinements not included, and why the conclusions are robust.} Two refinements are deliberately left out: electron heating and a Floquet-resolved dissipator~\cite{AkbariFloquet2026,AkbariFLT2026}. The absence of an SRPT rests on statements about the Hamiltonian alone: instantaneous positivity, Krein stability above $\omega_K$, the restriction of instabilities below $\omega_K$ to sum resonances, the gauge theorem for the Floquet mean field, and the Fock--Darwin high-frequency limit. Heating raises the electron temperature $T_e$ and changes parameters such as $\gamma$ and $D_{\rm eff}$, but it preserves minimal coupling and therefore positivity. 

Following Refs.~\cite{Kim2026,SqueezingDraft}, heating would raise the thresholds, shift the polaritons through $D_{\rm eff}(T_e)$, add thermal noise that degrades the squeezing, and clamp above-threshold growth; a quantitative treatment requires an energy-balance model extended to include the Faraday field. A Floquet generalized master equation~\cite{AkbariFloquet2026,AkbariFLT2026}, which can differ qualitatively from a dressed-state master equation even for weak dissipation~\cite{AkbariFLT2026}, would evaluate the bath at the quasienergy sidebands $\omega_k\pm n\omega_p$, include non-secular terms, and define the emitted flux consistently with the Floquet steady state. Its main effect would be on the off-resonant background and on fluxes near threshold; thresholds and resonance positions are set by the Hamiltonian and the resonant linewidths, which our dissipator already captures.

\emph{Nonlocality.} The low magnetoplasmon of this work is the $q\to0$ (dipolar, sample-size) collective mode of the c.m.: the lower Fock--Darwin branch, or edge magnetoplasmon, with wavevector $q\sim1/R$ set by the sample size $R$, in the local regime $qr_c\ll1$ where Kohn's theorem holds ($r_c$ is the Larmor radius). It should be distinguished from the finite-$q$ bulk magnetoplasmons of an infinite sheet, poles of the screened longitudinal response $2\epsilon+q\chi^L(q,\omega)=0$. At $qr_c\sim1$ these hybridize with Bernstein modes near $2\omega_{\rm cyc}$, which gives the nonlocal feature near $1.5$--$1.6\,\omega_{\rm cyc}$ found for near-field excitation~\cite{Greten2026} and seen in nanostructured cavities~\cite{Rajabali2021}. A uniform cavity field and a uniform $B(t)$ couple only to $q=0$, so the finite-$q$ modes belong to the relative motion and decouple here. In nanoslot or near-field geometries, where the cavity field varies on the scale of $r_c$, Kohn's theorem fails, and the modulation would acquire additional pair-creation channels, with polaritons paired with finite-$q$ magnetoplasmons and nonlocally brightened $2\omega_{\rm cyc}$ transitions.

\emph{Nonlinearity and genuinely quantum regimes.} Equation~\eqref{eq:HK} is linear, so its stability analysis is classical. Kane nonparabolicity turns the dressed mode into a Kerr oscillator with $|\chi_{\rm K}|/\omega_{\rm cav}\sim5\times10^{-9}$~\cite{SqueezingDraft,Kim2026}. Its mean field is still classical (a Duffing oscillator), but its fluctuations can be squeezed below the vacuum level~\cite{SqueezingDraft}. Non-Gaussian effects would require $|\chi_{\rm K}|\gtrsim\kappa$, few electrons, or Landau-level-resolved effects near integer filling; only in these regimes would the classical mapping genuinely fail.

\section{Conclusions}
\label{sec:conclusions}
Periodic driving does not circumvent the no-go theorem for Landau polaritons. Including the Faraday field and using Kohn's theorem, the modulated problem reduces exactly to a linear c.m.--cavity model whose Hamiltonian is positive at every instant. It is stable to all orders above the Krein frequency, and below it admits only pair-creation instabilities. Its high-frequency limit is a Fock--Darwin problem rather than an enhanced coupling, and a gauge theorem excludes a Floquet condensate for any band structure. The predicted transition traces to averaging a Landau-basis matrix element separately from the Hamiltonian, which produces an effective model that violates the TRK sum rule. More generally, a driven model must respect gauge invariance dynamically, including the induced electric field, and not merely satisfy TRK at each instant. The same rule, to modulate and average the minimal-coupling Hamiltonian, applies to any linear USC system, with magnetic coupling to nonlinear spins as the exception. 

What modulation does offer is a tunable dynamical Casimir source, yielding Faraday-driven polariton--magnetoplasmon pairs at low threshold, two-mode squeezed photon pairs at the polariton sum resonance, and photon fluxes of order $10^9$--$10^{10}$~s$^{-1}$. This makes magnetically modulated Landau polaritons a natural target for Floquet engineering of USC~\cite{AkbariPRL2025,AkbariFloquet2026}. Electron heating and Floquet-resolved dissipation would refine the thresholds, fluxes, and squeezing levels, but not our general conclusions, which rest on the Hamiltonian alone.

\begin{acknowledgments}
This work was supported by the Natural Sciences and Engineering Research Council of Canada (NSERC) (Discovery Grant and Quantum Alliance); the Canadian Foundation for Innovation (CFI); and Queen's University, Canada.
\end{acknowledgments}

\appendix

\section{Parameters and physical scales}
\label{app:params}

\begin{table}[h!]
\caption{Parameters (units of $\omega_0$ unless stated) and physical scales for $\omega_0/2\pi=0.5$~THz.}
\label{tab:params}
\begin{ruledtabular}
\begin{tabular}{lll}
Quantity & Symbol & Value\\
\hline
cavity frequency & $\omega_{\rm cav}$ & $\omega_0$\\
coupling; $A^2$ coefficient & $\Omega_0$; $D$ & 0.5; 0.25\\
c.m.\ restoring force & $\omega_d$ & 0.2 (0 in Fig.~\ref{fig:highfreq}a,b)\\
Drude rate & $\gamma$ & 0.01\\
quality factor & $Q=\omega_0/\kappa_0$ & 5, 20\\
Kane gap & $E_g^*=-1/\alpha_{\rm np}$ & 1.98~eV\\
nonparabolicity & $\zeta=\hbar\omega_0/E_g^*$ & $\approx10^{-3}$\\
Kerr shift ($B=0$) & $|\chi_{\rm K}|/\omega_{\rm cav}$ & $\approx5\times10^{-9}$\\
LO runaway & $U_p$; $n_c$ & 10--15~meV; $\sim10^6$\\
flux $10^{-3}\omega_0$ & & $3.1\times10^9$~s$^{-1}$\\
Faraday field$^{\rm a}$ & $E$ & $\sim10^9$~V/m\\
Faraday field$^{\rm b}$ & $E$ & $\sim10^6$~V/m\\
\end{tabular}
\end{ruledtabular}
\raggedright\footnotesize $^{\rm a}\Delta B=5$~T, $\omega_p/2\pi=20$~THz, $r=2.5~\mu$m. $^{\rm b}\Delta B=0.1$~T, $\omega_p/2\pi=1$~THz, $r=2.5~\mu$m.
\end{table}

Table~\ref{tab:params} lists the parameters used in the calculations, in units of $\omega_0$, and converts the main quantities to physical units for $\omega_0/2\pi=0.5$~THz, corresponding to $B_0\approx1.2$~T in GaAs.

\section{Numerical methods, tests of the Krein theorem, and further lossless results}
\label{app:numerics}

\emph{Floquet exponents and stability maps.}
Monodromy matrices are obtained by fourth-order Runge--Kutta integration of $\dot\Phi=J\mathcal H(t)\Phi$. Using $\tau=\omega_pt$, $d\Phi/d\tau=J\mathcal H(\tau)\Phi/\omega_p$ with $\mathcal H(\tau)$ common to all $\omega_p$, so grids in $(\varepsilon,\omega_p)$ are batched. The step satisfies $\Delta\tau\,\nu_{\max}/\omega_p\le0.03$, with $\nu_{\max}$ the largest instantaneous eigenvalue. Points with $\lambda>10^{-4}\omega_0$ (maps) or $10^{-5}\omega_0$ (line scans, $\Delta\omega_p=10^{-3}\omega_0$) are classified as unstable. In the code we use the dimensionless form $H_K=\tfrac12[(P_x+wY/2-gx_a)^2+(P_y-wX/2)^2]+\tfrac12\omega_d^2(X^2+Y^2)+\tfrac12\omega_{\rm cav}(x_a^2+p_a^2)$ with $w=1+\varepsilon\cos\omega_pt$, $g^2=4D$, and $\omega_0=1$.

\begin{figure}[b]
\centering\includegraphics[width=\columnwidth]{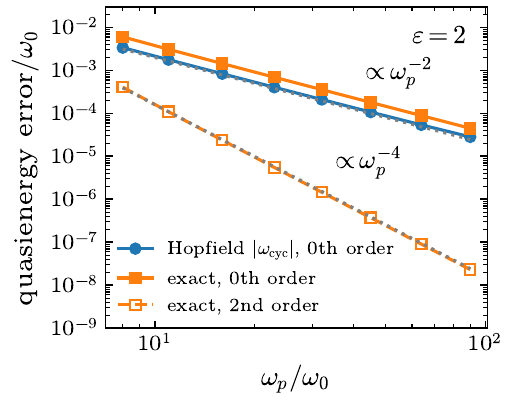}
\caption{Error of the high-frequency expansion against exact quasienergies ($\varepsilon=2$): zeroth order for the Hopfield model with $|\omega_{\rm cyc}|$ (blue) and the exact model (orange), and second order for the exact model (orange, open). Dotted: $\omega_p^{-2}$ and $\omega_p^{-4}$.}
\label{fig:conv}
\end{figure}

\begin{figure*}[t]
\centering\includegraphics[width=\textwidth]{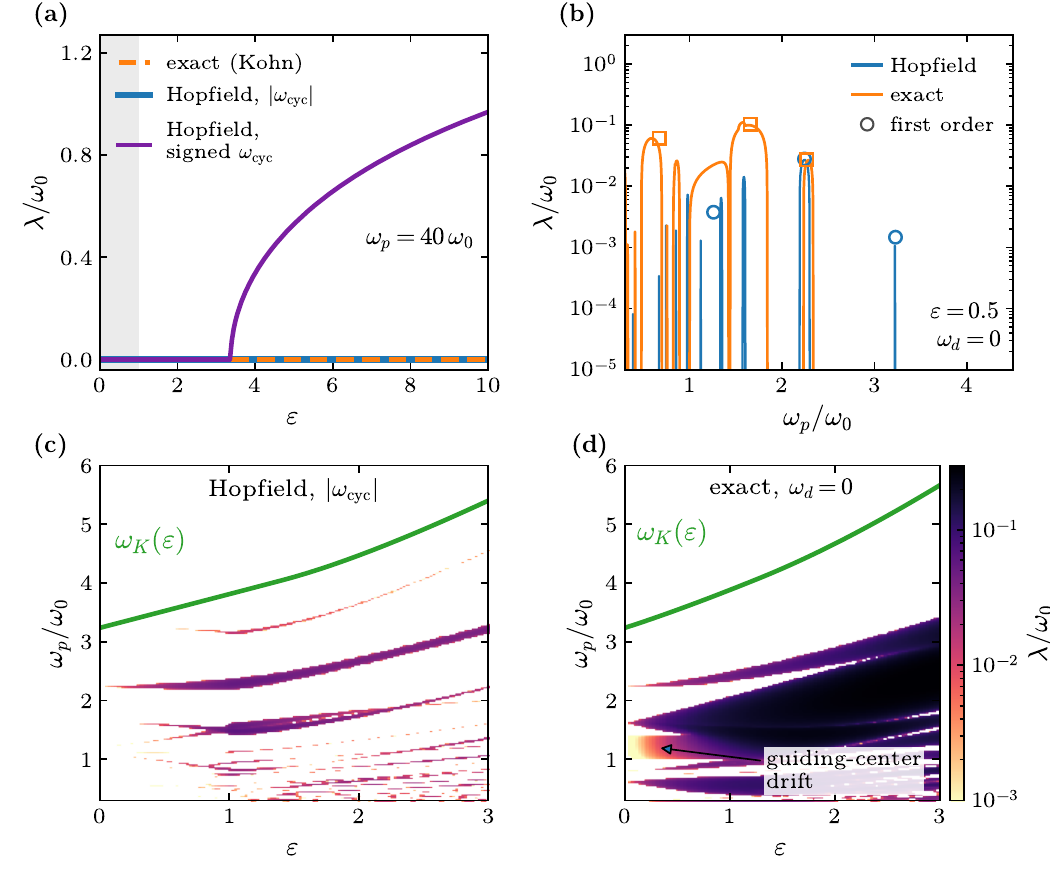}
\caption{Lossless Floquet exponents for an infinite sample (colors as in Fig.~\ref{fig:highfreq}: purple, signed Hopfield; blue, Hopfield with $|\omega_{\rm cyc}|$; orange, exact). (a)~$\omega_p=40\,\omega_0$. (b)~$\varepsilon=0.5$, $\omega_d=0$; symbols: first-order (RWA) estimates $|K^{(1)}_{ij}|$. (c,d)~$(\varepsilon,\omega_p)$ maps for the Hopfield model and the exact model with $\omega_d=0$, with the Krein frequency $\omega_K(\varepsilon)$ (green). The weak band in (d) near $\omega_p\approx1.2\,\omega_0$ is the guiding-center drift of an infinite electron gas.}
\label{fig:lossless}
\end{figure*}

\begin{figure*}[t]
\centering\includegraphics[width=\textwidth]{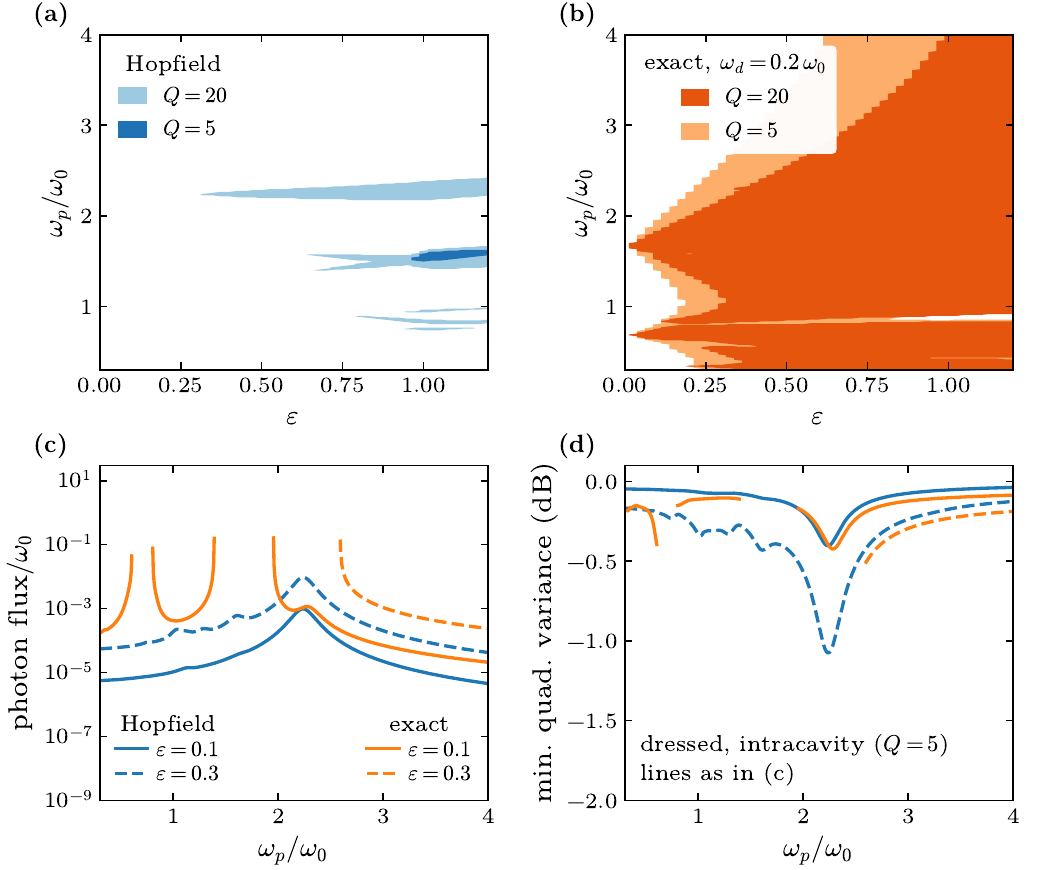}
\caption{Loss ($\gamma=0.01\,\omega_0$, $\hat\Pi_c=\aop$). (a)~Hopfield model above threshold (light blue: $Q=20$; dark blue: $Q=5$). (b)~Exact model with $\omega_d=0.2\,\omega_0$ (light orange: $Q=5$; dark orange: $Q=20$, lying almost entirely inside the $Q=5$ region). (c)~Emitted photon flux and (d)~minimum intracavity quadrature variance in dressed-mode coordinates for $Q=5$; blue: Hopfield, orange: exact; solid: $\varepsilon=0.1$, dashed: $\varepsilon=0.3$. Gaps mark points above threshold.}
\label{fig:loss}
\end{figure*}

\emph{High-frequency expansion.} For quadratic forms, $[\hat H_A,\hat H_B]=\hat H_C$ with $C=i(AJB-BJA)$. The van Vleck Hamiltonian~\cite{Eckardt2015},
\begin{equation}
\begin{split}
H_{\rm eff}={}&H_0+\sum_{m>0}\frac{[H_m,H_{-m}]}{m\omega_p}+\sum_{m\neq0}\frac{[[H_{-m},H_0],H_m]}{2m^2\omega_p^2}\\
&+\sum_{m\neq0}\sum_{m'\neq0,m}\frac{[[H_{-m'},H_{m'-m}],H_m]}{3mm'\omega_p^2},
\end{split}
\end{equation}
loses its $O(\omega_p^{-1})$ term for the cosine drive, $H(t)=H(-t)$. Figure~\ref{fig:conv} shows that the zeroth-order error scales as $\omega_p^{-2}$ and the second-order error as $\omega_p^{-4}$.

\emph{Krein frequency and tests of Theorem~\ref{thm:krein}.} To evaluate Eq.~\eqref{eq:wK}, we optimize over diagonal symplectic scalings, i.e., independent rescalings of each conjugate pair, with multiple starting points. For $\Omega_0=0.5\,\omega_0$, $\omega_K=40\,\omega_0$ is reached only at $\varepsilon\approx31$ (Hopfield, $|\omega_{\rm cyc}|$) or 28 (Kohn), and $\omega_K=13.7\,\omega_0$ at $\varepsilon=10$. In all maps, no unstable point lies above $\omega_K$. To test Theorem~\ref{thm:krein}, we generated 60 random smooth positive systems with $\int\lambda_{\max}\,dt=0.98\pi$; none was unstable, whereas 50 of 60 were unstable at $2.5\pi$. A second test used 120 strongly pulsed systems, $\mathcal H(t)=e^{a\cos(t-\phi)}L(t)^{\rm T}L(t)$ with random trigonometric $L(t)$, $a\le6$, and $\sup\lambda_{\max}/\langle\lambda_{\max}\rangle$ up to 8, integrated with a fourth-order Magnus propagator. None was unstable at $\int\lambda_{\max}\,dt=0.97\pi$, whereas 29 of 40 were unstable at $2\pi$.

\emph{Further lossless results.} Figure~\ref{fig:lossless} shows results for an infinite sample ($\omega_d=0$). The first-order rates of Eq.~\eqref{eq:rates} agree with the exact peaks to within about 10\% [Fig.~\ref{fig:lossless}(b)]. For $\omega_d=0$, a weak band near $\omega_p\approx1.2\,\omega_0$, with $\lambda\propto\varepsilon^2$ and real multipliers near 1, is an anti-trapping of the guiding-center zero mode by a second-order effective potential [Fig.~\ref{fig:lossless}(d)]. It is absent for $\omega_d=0.2\,\omega_0$ ($\lambda<2\times10^{-4}$ for $\varepsilon\le0.3$).

\section{Open-system details and robustness}
\label{app:open}

\emph{Level of approximation.} The Hamiltonian part of Eq.~\eqref{eq:lyap} is exact, including all counter-rotating and Faraday terms. The dissipator makes three simplifications: (i)~the jump operators are built from the static (undriven) dressed modes; (ii)~it is secular, with one jump operator per dressed mode; and (iii)~it is not Floquet resolved. To test (ii), we replaced the cavity part by a single non-secular channel, $\hat L=\sqrt{\kappa_0}\sum_ku_k\hat c_k$. The thresholds are unchanged (minimum $\varepsilon_{\rm th}=0.05$ for the exact model at $Q=5$), and the unstable area for $\varepsilon\le0.6$ decreases from 61\% to 52\%. The fluxes agree to within 10\% for the Hopfield model and to within a factor of 3 for the exact model near threshold, where $\Phi_{\rm ph}\propto1/|\lambda|$ is most sensitive.

\emph{Definition of the flux.} Under driving, the number of photons depends on which modes are used. Equation~\eqref{eq:flux} uses the undriven dressed modes with separate output channels, consistent with the dissipator. We compared it with a single cavity channel, $\kappa_0\overline{\langle\hat x^{+\dagger}\hat x^+\rangle}$ with $\hat x^+=\sum_ku_k\hat c_k$, and with the instantaneous dressed modes of $\mathcal H(t)$. At the Hopfield $\omega_-+\omega_+$ resonance all three agree to within 2\% ($9.75$, $9.75$, and $9.79\times10^{-4}\omega_0$ at $\varepsilon=0.1$). Near the Faraday tongues of the exact model they agree to within a factor of 2. Far off resonance they differ by up to a factor of 6 (exact model, $\omega_p=4\,\omega_0$, $2.1\times10^{-5}$ versus $1.3\times10^{-4}\omega_0$). The off-resonant background therefore mixes genuine emission with adiabatic dressing of the vacuum, and only a Floquet-resolved treatment separates them. The resonant features and thresholds are robust.

\emph{Further results.} Figure~\ref{fig:loss} shows the threshold maps of both models separately, together with the intracavity squeezing in dressed-mode coordinates. The intracavity squeezing is $\approx1.1$~dB at $\varepsilon=0.3$, below the 3~dB intracavity bound; the output squeezing of Fig.~\ref{fig:squeezing} is much larger near threshold, as for parametric amplifiers.

\bigskip

\clearpage
\bibliography{prx}

@misc{PerezGonzalez2026,
  author        = {P{\'e}rez-Gonz{\'a}lez, B. and Kohler, S. and Benito, M.},
  title         = {Floquet physics from quantized light--matter interaction: geometric phases, gauge consistency, and entanglement},
  year          = {2026},
  eprint        = {2609.21741},
  archivePrefix = {arXiv},
  url           = {https://arxiv.org/abs/2609.21741}
}

@misc{AkbariFLT2026,
  author        = {Akbari, K. and Hughes, S.},
  title         = {{F}loquet--{L}iouville theory for strongly driven open quantum systems},
  year          = {2026},
  eprint        = {2608.14966},
  archivePrefix = {arXiv},
  url           = {https://arxiv.org/abs/2608.14966}
}

@misc{Kim2026,
  author  = {Kim, D. and Kulkarni, K. M. and Sharma, V. and Lee, D. and Lee, G. and Kim, S. and Grumm, J. and Liang, S. and Xu, H. and Tay, F. and Baydin, A. and Bamba, M. and Knorr, A. and Stanton, C. J. and Manfra, M. J. and Seo, M. and Hughes, S. and Kono, J.},
  title   = {Nonlinear Diamagnetic Interactions in Ultrastrongly Coupled {2D} Electrons},
  year    = {2026},
  eprint  = {2608.20578},
  archivePrefix = {arXiv},
  doi     = {10.48550/arXiv.2608.20578},
  url     = {https://arxiv.org/abs/2608.20578}
}

@article{Gustin2023,
  author  = {Gustin, C. and Franke, S. and Hughes, S.},
  title   = {Gauge-invariant theory of truncated quantum light-matter interactions in arbitrary media},
  journal = {Phys. Rev. A},
  volume  = {107},
  pages   = {013722},
  year    = {2023},
  doi     = {10.1103/PhysRevA.107.013722}
}

@article{Garziano2020,
  author  = {Garziano, L. and Settineri, A. and Di Stefano, O. and Savasta, S. and Nori, F.},
  title   = {Gauge invariance of the {D}icke and {H}opfield models},
  journal = {Phys. Rev. A},
  volume  = {102},
  pages   = {023718},
  year    = {2020},
  doi     = {10.1103/PhysRevA.102.023718}
}

@article{DeBernardis2018,
  author  = {De Bernardis, D. and Pilar, P. and Jaako, T. and De Liberato, S. and Rabl, P.},
  title   = {Breakdown of gauge invariance in ultrastrong-coupling cavity {QED}},
  journal = {Phys. Rev. A},
  volume  = {98},
  pages   = {053819},
  year    = {2018},
  doi     = {10.1103/PhysRevA.98.053819}
}

@article{Hughes2024,
  author  = {Hughes, S. and Gustin, C. and Nori, F.},
  title   = {Reconciling quantum and classical spectral theories of ultrastrong coupling: role of cavity bath coupling and gauge corrections},
  journal = {Opt. Quantum},
  volume  = {2},
  pages   = {133},
  year    = {2024},
  doi     = {10.1364/OPTICAQ.519395}
}

@article{LSA2026,
  author  = {Gustin, C. and Ren, J. and Franke, S. and Hughes, S.},
  title   = {Dissipation in the broadband and ultrastrong coupling regimes of cavity quantum electrodynamics: an ab initio quantized quasinormal mode approach},
  journal = {Light Sci. Appl.},
  volume  = {15},
  pages   = {364},
  year    = {2026},
  doi     = {10.1038/s41377-026-02406-2}
}

@article{GustinPRL2025,
  author  = {Gustin, C. and Ren, J. and Hughes, S.},
  title   = {What is the spectral density of the reservoir for a lossy quantized cavity?},
  journal = {Phys. Rev. Lett.},
  volume  = {134},
  pages   = {123601},
  year    = {2025},
  doi     = {10.1103/PhysRevLett.134.123601}
}

@article{Lamberto2025,
  author  = {Lamberto, D. and Di Stefano, O. and Hughes, S. and Nori, F. and Savasta, S.},
  title   = {Quantum phase transitions in many-dipole light-matter systems},
  journal = {Phys. Rev. Res.},
  volume  = {7},
  pages   = {013271},
  year    = {2025},
  doi     = {10.1103/PhysRevResearch.7.013271}
}

@article{DiStefano2019,
  author  = {Di Stefano, O. and Settineri, A. and Macr{\`i}, V. and Garziano, L. and Stassi, R. and Savasta, S. and Nori, F.},
  title   = {Resolution of gauge ambiguities in ultrastrong-coupling cavity quantum electrodynamics},
  journal = {Nat. Phys.},
  volume  = {15},
  pages   = {803},
  year    = {2019},
  doi     = {10.1038/s41567-019-0534-4}
}

@article{Hagenmuller2010,
  author  = {Hagenm{\"u}ller, D. and De Liberato, S. and Ciuti, C.},
  title   = {Ultrastrong coupling between a cavity resonator and the cyclotron transition of a two-dimensional electron gas in the case of an integer filling factor},
  journal = {Phys. Rev. B},
  volume  = {81},
  pages   = {235303},
  year    = {2010},
  doi     = {10.1103/PhysRevB.81.235303}
}

@article{Scalari2012,
  author  = {Scalari, G. and Maissen, C. and Tur{\v{c}}inkov{\'a}, D. and Hagenm{\"u}ller, D. and De Liberato, S. and Ciuti, C. and Reichl, C. and Schuh, D. and Wegscheider, W. and Beck, M. and Faist, J.},
  title   = {Ultrastrong coupling of the cyclotron transition of a {2D} electron gas to a {THz} metamaterial},
  journal = {Science},
  volume  = {335},
  pages   = {1323},
  year    = {2012},
  doi     = {10.1126/science.1216022}
}

@article{Li2018,
  author  = {Li, X. and Bamba, M. and Zhang, Q. and Fallahi, S. and Gardner, G. C. and Gao, W. and Lou, M. and Yoshioka, K. and Manfra, M. J. and Kono, J.},
  title   = {Vacuum {B}loch--{S}iegert shift in {L}andau polaritons with ultra-high cooperativity},
  journal = {Nat. Photon.},
  volume  = {12},
  pages   = {324},
  year    = {2018},
  doi     = {10.1038/s41566-018-0153-0}
}

@article{Kohn1961,
  author  = {Kohn, W.},
  title   = {Cyclotron resonance and de {H}aas--van {A}lphen oscillations of an interacting electron gas},
  journal = {Phys. Rev.},
  volume  = {123},
  pages   = {1242},
  year    = {1961},
  doi     = {10.1103/PhysRev.123.1242}
}

@article{Kane1957,
  author  = {Kane, E. O.},
  title   = {Band structure of indium antimonide},
  journal = {J. Phys. Chem. Solids},
  volume  = {1},
  pages   = {249},
  year    = {1957},
  doi     = {10.1016/0022-3697(57)90013-6}
}

@article{Ciuti2005,
  author  = {Ciuti, C. and Bastard, G. and Carusotto, I.},
  title   = {Quantum vacuum properties of the intersubband cavity polariton field},
  journal = {Phys. Rev. B},
  volume  = {72},
  pages   = {115303},
  year    = {2005},
  doi     = {10.1103/PhysRevB.72.115303}
}

@article{Andolina2019,
  author  = {Andolina, G. M. and Pellegrino, F. M. D. and Giovannetti, V. and MacDonald, A. H. and Polini, M.},
  title   = {Cavity quantum electrodynamics of strongly correlated electron systems: A no-go theorem for photon condensation},
  journal = {Phys. Rev. B},
  volume  = {100},
  pages   = {121109},
  year    = {2019},
  doi     = {10.1103/PhysRevB.100.121109}
}

@misc{AkbariFloquet2026,
  author  = {Akbari, K. and Nori, F. and Hughes, S.},
  title   = {Floquet quasienergy-resolved dissipation, dynamics, and spectroscopy in ultrastrong cavity-{QED}},
  year    = {2026},
  eprint  = {2606.31108},
  archivePrefix = {arXiv},
  note    = {arXiv:2606.31108},
  url     = {https://arxiv.org/abs/2606.31108}
}

@article{HolsteinPrimakoff1940,
  author  = {Holstein, T. and Primakoff, H.},
  title   = {Field dependence of the intrinsic domain magnetization of a ferromagnet},
  journal = {Phys. Rev.},
  volume  = {58},
  pages   = {1098},
  year    = {1940},
  doi     = {10.1103/PhysRev.58.1098}
}

@article{Gunter2009,
  author  = {G{\"u}nter, G. and Anappara, A. A. and Hees, J. and Sell, A. and Biasiol, G. and Sorba, L. and De Liberato, S. and Ciuti, C. and Tredicucci, A. and Leitenstorfer, A. and Huber, R.},
  title   = {Sub-cycle switch-on of ultrastrong light--matter interaction},
  journal = {Nature},
  volume  = {458},
  pages   = {178},
  year    = {2009},
  doi     = {10.1038/nature07838}
}

@article{DeLiberato2007,
  author  = {De Liberato, S. and Ciuti, C. and Carusotto, I.},
  title   = {Quantum vacuum radiation spectra from a semiconductor microcavity with a time-modulated vacuum {R}abi frequency},
  journal = {Phys. Rev. Lett.},
  volume  = {98},
  pages   = {103602},
  year    = {2007},
  doi     = {10.1103/PhysRevLett.98.103602}
}

@book{HardyLittlewoodPolya1952,
  author    = {Hardy, G. H. and Littlewood, J. E. and P{\'o}lya, G.},
  title     = {Inequalities},
  edition   = {2nd},
  publisher = {Cambridge University Press},
  address   = {Cambridge},
  year      = {1952}
}

@article{Riek2015,
  author  = {Riek, C. and Seletskiy, D. V. and Moskalenko, A. S. and Schmidt, J. F. and Krauspe, P. and Eckart, S. and Eggert, S. and Burkard, G. and Leitenstorfer, A.},
  title   = {Direct sampling of electric-field vacuum fluctuations},
  journal = {Science},
  volume  = {350},
  pages   = {420},
  year    = {2015},
  doi     = {10.1126/science.aac9788}
}

@article{BeneaChelmus2019,
  author  = {Benea-Chelmus, I.-C. and Settembrini, F. F. and Scalari, G. and Faist, J.},
  title   = {Electric field correlation measurements on the electromagnetic vacuum state},
  journal = {Nature},
  volume  = {568},
  pages   = {202},
  year    = {2019},
  doi     = {10.1038/s41586-019-1083-9}
}

@article{Zhang2016,
  author  = {Zhang, Q. and Lou, M. and Li, X. and Reno, J. L. and Pan, W. and Watson, J. D. and Manfra, M. J. and Kono, J.},
  title   = {Collective non-perturbative coupling of {2D} electrons with high-quality-factor terahertz cavity photons},
  journal = {Nat. Phys.},
  volume  = {12},
  pages   = {1005},
  year    = {2016},
  doi     = {10.1038/nphys3850}
}

@article{Rajabali2021,
  author  = {Rajabali, S. and Cortese, E. and Beck, M. and De Liberato, S. and Faist, J. and Scalari, G.},
  title   = {Polaritonic nonlocality in light--matter interaction},
  journal = {Nat. Photon.},
  volume  = {15},
  pages   = {690},
  year    = {2021},
  doi     = {10.1038/s41566-021-00854-3}
}

@article{FriskKockum2019,
  author  = {Frisk Kockum, A. and Miranowicz, A. and De Liberato, S. and Savasta, S. and Nori, F.},
  title   = {Ultrastrong coupling between light and matter},
  journal = {Nat. Rev. Phys.},
  volume  = {1},
  pages   = {19},
  year    = {2019},
  doi     = {10.1038/s42254-018-0006-2}
}

@article{FornDiaz2019,
  author  = {Forn-D\'{\i}az, P. and Lamata, L. and Rico, E. and Kono, J. and Solano, E.},
  title   = {Ultrastrong coupling regimes of light-matter interaction},
  journal = {Rev. Mod. Phys.},
  volume  = {91},
  pages   = {025005},
  year    = {2019},
  doi     = {10.1103/RevModPhys.91.025005}
}

@article{Niemczyk2010,
  author  = {Niemczyk, T. and Deppe, F. and Huebl, H. and Menzel, E. P. and Hocke, F. and Schwarz, M. J. and Garcia-Ripoll, J. J. and Zueco, D. and H\"{u}mmer, T. and Solano, E. and Marx, A. and Gross, R.},
  title   = {Circuit quantum electrodynamics in the ultrastrong-coupling regime},
  journal = {Nat. Phys.},
  volume  = {6},
  pages   = {772},
  year    = {2010},
  doi     = {10.1038/nphys1730}
}

@article{Yoshihara2017,
  author  = {Yoshihara, F. and Fuse, T. and Ashhab, S. and Kakuyanagi, K. and Saito, S. and Semba, K.},
  title   = {Superconducting qubit--oscillator circuit beyond the ultrastrong-coupling regime},
  journal = {Nat. Phys.},
  volume  = {13},
  pages   = {44},
  year    = {2017},
  doi     = {10.1038/nphys3906}
}

@article{Bayer2017,
  author  = {Bayer, A. and Pozimski, M. and Schambeck, S. and Schuh, D. and Huber, R. and Bougeard, D. and Lange, C.},
  title   = {Terahertz light--matter interaction beyond unity coupling strength},
  journal = {Nano Lett.},
  volume  = {17},
  pages   = {6340},
  year    = {2017},
  doi     = {10.1021/acs.nanolett.7b03103}
}

@article{Hopfield1958,
  author  = {Hopfield, J. J.},
  title   = {Theory of the contribution of excitons to the complex dielectric constant of crystals},
  journal = {Phys. Rev.},
  volume  = {112},
  pages   = {1555},
  year    = {1958},
  doi     = {10.1103/PhysRev.112.1555}
}

@book{LandauMechanics,
  author    = {Landau, L. D. and Lifshitz, E. M.},
  title     = {Mechanics},
  edition   = {3rd},
  publisher = {Pergamon},
  address   = {Oxford},
  year      = {1976}
}

@misc{Wu2026,
  author        = {Wu, W.-H. and Tay, F. and Che, M. and Baydin, A. and Kono, J. and Hagenm{\"u}ller, D.},
  title         = {Floquet engineering of a quasiequilibrium superradiant phase transition in {L}andau polaritons},
  year          = {2026},
  eprint        = {2604.08635v1},
  archivePrefix = {arXiv},
  url           = {https://arxiv.org/abs/2604.08635v1}
}

@article{Rzazewski1975,
  author = {Rz{\c{a}}{\.z}ewski, K. and W{\'o}dkiewicz, K. and {\.Z}akowicz, W.},
  title  = {Phase transitions, two-level atoms, and the {$A^2$} term},
  journal = {Phys. Rev. Lett.}, volume = {35}, pages = {432}, year = {1975}, doi = {10.1103/PhysRevLett.35.432}
}

@article{Nataf2010,
  author = {Nataf, P. and Ciuti, C.},
  title  = {No-go theorem for superradiant quantum phase transitions in cavity {QED} and counter-example in circuit {QED}},
  journal = {Nat. Commun.}, volume = {1}, pages = {72}, year = {2010}, doi = {10.1038/ncomms1069}
}

@article{Eckardt2015,
  author = {Eckardt, A. and Anisimovas, E.},
  title  = {High-frequency approximation for periodically driven quantum systems from a {F}loquet-space perspective},
  journal = {New J. Phys.}, volume = {17}, pages = {093039}, year = {2015}, doi = {10.1088/1367-2630/17/9/093039}
}

@article{AkbariPRL2025,
  author = {Akbari, K. and Nori, F. and Hughes, S.},
  title  = {Floquet engineering the quantum {R}abi model in the ultrastrong coupling regime},
  journal = {Phys. Rev. Lett.}, volume = {134}, pages = {063602}, year = {2025}, doi = {10.1103/PhysRevLett.134.063602}
}

@article{ChangPRA2025,
  title  = {Direct space-time modeling of mechanically dressed dipole-dipole interactions with electromagnetically coupled oscillating dipoles},
  author = {Chang, Yi-Ming and Akbari, Kamran and Filipovich, Matthew and Hughes, Stephen},
  journal = {Phys. Rev. A}, volume = {112}, pages = {033530}, year = {2025}, doi = {10.1103/bmm8-9pn6}
}

@article{Bamba2014,
  author = {Bamba, M. and Ogawa, T.},
  title  = {Stability of polarizable materials against superradiant phase transition},
  journal = {Phys. Rev. A}, volume = {90}, pages = {063825}, year = {2014}, doi = {10.1103/PhysRevA.90.063825}
}

@article{PyCharge,
  author = {Filipovich, M. J. and Hughes, S.},
  title  = {{PyCharge}: An open-source {P}ython package for self-consistent electrodynamics simulations of {L}orentz oscillators and moving point charges},
  journal = {Comput. Phys. Commun.}, volume = {274}, pages = {108291}, year = {2022}, doi = {10.1016/j.cpc.2022.108291}
}

@misc{SqueezingDraft,
  author = {},
  title  = {Squeezed terahertz light from nonlinear diamagnetism in ultrastrongly coupled two-dimensional electrons},
  note   = {in preparation}, year = {2026}
}

@book{YakubovichStarzhinskii1975, author = {Yakubovich, V. A. and Starzhinskii, V. M.}, title = {Linear Differential Equations with Periodic Coefficients}, publisher = {Wiley}, address = {New York}, year = {1975}}

@incollection{Krein1955, author = {Krein, M. G.}, title = {The basic propositions of the theory of $\lambda$-zones of stability of a canonical system of linear differential equations with periodic coefficients}, booktitle = {In Memory of A. A. Andronov}, publisher = {Izdat. Akad. Nauk SSSR}, address = {Moscow}, year = {1955}, pages = {413--498}, note = {{E}nglish transl.: Amer. Math. Soc. Transl. (2) \textbf{120}, 1 (1983) }}

@article{GelfandLidskii1955, author = {Gel'fand, I. M. and Lidskii, V. B.}, title = {On the structure of the regions of stability of linear canonical systems of differential equations with periodic coefficients}, journal = {Usp. Mat. Nauk}, volume = {10}, number = {1}, pages = {3}, year = {1955}, note = {{E}nglish transl.: Amer. Math. Soc. Transl. (2) \textbf{8}, 143 (1958) }}

@article{Moser1958, author = {Moser, J.}, title = {New aspects in the theory of stability of {H}amiltonian systems}, journal = {Commun. Pure Appl. Math.}, volume = {11}, pages = {81}, year = {1958}, doi = {10.1002/cpa.3160110105}}

@book{Ekeland1990, author = {Ekeland, I.}, title = {Convexity Methods in Hamiltonian Mechanics}, publisher = {Springer}, address = {Berlin}, year = {1990}, doi = {10.1007/978-3-642-74331-3}}

@book{Arnold1989, author = {Arnold, V. I.}, title = {Mathematical Methods of Classical Mechanics}, edition = {2nd}, publisher = {Springer}, address = {New York}, year = {1989}, doi = {10.1007/978-1-4757-2063-1}}

@article{Kovacic2018, author = {Kovacic, I. and Rand, R. and Sah, S. M.}, title = {Mathieu's equation and its generalizations: Overview of stability charts and their features}, journal = {Appl. Mech. Rev.}, volume = {70}, pages = {020802}, year = {2018}, doi = {10.1115/1.4039144}}

@book{ZhangSchur2005, editor = {Zhang, F.}, title = {The {S}chur Complement and Its Applications}, publisher = {Springer}, address = {New York}, year = {2005}, doi = {10.1007/b105056}}

@book{HornJohnson2013, author = {Horn, R. A. and Johnson, C. R.}, title = {Matrix Analysis}, edition = {2nd}, publisher = {Cambridge University Press}, address = {Cambridge}, year = {2013}, doi = {10.1017/CBO9781139020411}}

@article{Brey1989, author = {Brey, L. and Johnson, N. F. and Halperin, B. I.}, title = {Optical and magneto-optical absorption in parabolic quantum wells}, journal = {Phys. Rev. B}, volume = {40}, pages = {10647}, year = {1989}, doi = {10.1103/PhysRevB.40.10647}}

@article{MaksymChakraborty1990, author = {Maksym, P. A. and Chakraborty, T.}, title = {Quantum dots in a magnetic field: Role of electron-electron interactions}, journal = {Phys. Rev. Lett.}, volume = {65}, pages = {108}, year = {1990}, doi = {10.1103/PhysRevLett.65.108}}

@article{Peeters1990, author = {Peeters, F. M.}, title = {Magneto-optics in parabolic quantum dots}, journal = {Phys. Rev. B}, volume = {42}, pages = {1486}, year = {1990}, doi = {10.1103/PhysRevB.42.1486}}

@article{Dobson1994, author = {Dobson, J. F.}, title = {Harmonic-potential theorem: Implications for approximate many-body theories}, journal = {Phys. Rev. Lett.}, volume = {73}, pages = {2244}, year = {1994}, doi = {10.1103/PhysRevLett.73.2244}}

@article{Rokaj2019, author = {Rokaj, V. and Penz, M. and Sentef, M. A. and Ruggenthaler, M. and Rubio, A.}, title = {Quantum electrodynamical {B}loch theory with homogeneous magnetic fields}, journal = {Phys. Rev. Lett.}, volume = {123}, pages = {047202}, year = {2019}, doi = {10.1103/PhysRevLett.123.047202}}

@article{Rokaj2023, author = {Rokaj, V. and Wang, J. and Sous, J. and Penz, M. and Ruggenthaler, M. and Rubio, A.}, title = {Weakened topological protection of the quantum {H}all effect in a cavity}, journal = {Phys. Rev. Lett.}, volume = {131}, pages = {196602}, year = {2023}, doi = {10.1103/PhysRevLett.131.196602}}

@article{Fock1928, author = {Fock, V.}, title = {Bemerkung zur {Q}uantelung des harmonischen {O}szillators im {M}agnetfeld}, journal = {Z. Phys.}, volume = {47}, pages = {446}, year = {1928}, doi = {10.1007/BF01390750}}

@article{Darwin1931, author = {Darwin, C. G.}, title = {The diamagnetism of the free electron}, journal = {Proc. Cambridge Philos. Soc.}, volume = {27}, pages = {86}, year = {1931}, url = {https://www.cambridge.org/core/journals/mathematical-proceedings-of-the-cambridge-philosophical-society/article/diamagnetism-of-the-free-electron/992F1A901D9E9B1B0680F7CB0BF01381}}

@article{Kapitza1951, author = {Kapitza, P. L.}, title = {Dynamic stability of a pendulum when its point of suspension vibrates}, journal = {Zh. Eksp. Teor. Fiz.}, volume = {21}, pages = {588}, year = {1951}}

@article{Moore1970, author = {Moore, G. T.}, title = {Quantum theory of the electromagnetic field in a variable-length one-dimensional cavity}, journal = {J. Math. Phys.}, volume = {11}, pages = {2679}, year = {1970}, doi = {10.1063/1.1665432}}

@article{FullingDavies1976, author = {Fulling, S. A. and Davies, P. C. W.}, title = {Radiation from a moving mirror in two dimensional space-time: conformal anomaly}, journal = {Proc. R. Soc. Lond. A}, volume = {348}, pages = {393}, year = {1976}, doi = {10.1098/rspa.1976.0045}}

@article{Dodonov2010, author = {Dodonov, V. V.}, title = {Current status of the dynamical {C}asimir effect}, journal = {Phys. Scr.}, volume = {82}, pages = {038105}, year = {2010}, doi = {10.1088/0031-8949/82/03/038105}}

@article{Nation2012, author = {Nation, P. D. and Johansson, J. R. and Blencowe, M. P. and Nori, F.}, title = {Colloquium: Stimulating uncertainty: Amplifying the quantum vacuum with superconducting circuits}, journal = {Rev. Mod. Phys.}, volume = {84}, pages = {1}, year = {2012}, doi = {10.1103/RevModPhys.84.1}}

@article{Wilson2011, author = {Wilson, C. M. and Johansson, G. and Pourkabirian, A. and Simoen, M. and Johansson, J. R. and Duty, T. and Nori, F. and Delsing, P.}, title = {Observation of the dynamical {C}asimir effect in a superconducting circuit}, journal = {Nature}, volume = {479}, pages = {376}, year = {2011}, doi = {10.1038/nature10561}}

@article{Lahteenmaki2013, author = {L{\"a}hteenm{\"a}ki, P. and Paraoanu, G. S. and Hassel, J. and Hakonen, P. J.}, title = {Dynamical {C}asimir effect in a {J}osephson metamaterial}, journal = {Proc. Natl. Acad. Sci. USA}, volume = {110}, pages = {4234}, year = {2013}, doi = {10.1073/pnas.1212705110}}

@article{Hagenmuller2016, author = {Hagenm{\"u}ller, D.}, title = {All-optical dynamical {C}asimir effect in a three-dimensional terahertz photonic band gap}, journal = {Phys. Rev. B}, volume = {93}, pages = {235309}, year = {2016}, doi = {10.1103/PhysRevB.93.235309}}

@article{Galiffi2022, author = {Galiffi, E. and Tirole, R. and Yin, S. and Li, H. and Vezzoli, S. and Huidobro, P. A. and Silveirinha, M. G. and Sapienza, R. and Al{\`u}, A. and Pendry, J. B.}, title = {Photonics of time-varying media}, journal = {Adv. Photonics}, volume = {4}, pages = {014002}, year = {2022}, doi = {10.1117/1.AP.4.1.014002}}

@article{Lustig2018, author = {Lustig, E. and Sharabi, Y. and Segev, M.}, title = {Topological aspects of photonic time crystals}, journal = {Optica}, volume = {5}, pages = {1390}, year = {2018}, doi = {10.1364/OPTICA.5.001390}}

@article{Andolina2020, author = {Andolina, G. M. and Pellegrino, F. M. D. and Giovannetti, V. and MacDonald, A. H. and Polini, M.}, title = {Theory of photon condensation in a spatially varying electromagnetic field}, journal = {Phys. Rev. B}, volume = {102}, pages = {125137}, year = {2020}, doi = {10.1103/PhysRevB.102.125137}}

@article{Appugliese2022, author = {Appugliese, F. and Enkner, J. and Paravicini-Bagliani, G. L. and Beck, M. and Reichl, C. and Wegscheider, W. and Scalari, G. and Ciuti, C. and Faist, J.}, title = {Breakdown of topological protection by cavity vacuum fields in the integer quantum {H}all effect}, journal = {Science}, volume = {375}, pages = {1030}, year = {2022}}

@article{BeaudoinGambettaBlais2011, author = {Beaudoin, F. and Gambetta, J. M. and Blais, A.}, title = {Dissipation and ultrastrong coupling in circuit {QED}}, journal = {Phys. Rev. A}, volume = {84}, pages = {043832}, year = {2011}, doi = {10.1103/PhysRevA.84.043832}}

@article{Muniain2025, author = {Muniain, U. and Aizpurua, J. and Hillenbrand, R. and Mart{\'\i}n-Moreno, L. and Esteban, R.}, title = {Description of ultrastrong light-matter interaction through coupled harmonic oscillator models and their connection with cavity-{QED} {H}amiltonians}, journal = {Nanophotonics}, year = {2025}, doi = {10.1515/nanoph-2024-0528}}

@article{Andolina2022, author = {Andolina, G. M. and Pellegrino, F. M. D. and Mercurio, A. and Di Stefano, O. and Polini, M. and Savasta, S.}, title = {A non-perturbative no-go theorem for photon condensation in approximate models}, journal = {Eur. Phys. J. Plus}, volume = {137}, pages = {1348}, year = {2022}, doi = {10.1140/epjp/s13360-022-03571-0}}

@article{Aspelmeyer2014, author = {Aspelmeyer, M. and Kippenberg, T. J. and Marquardt, F.}, title = {Cavity optomechanics}, journal = {Rev. Mod. Phys.}, volume = {86}, pages = {1391}, year = {2014}, doi = {10.1103/RevModPhys.86.1391}}

@misc{Greten2026, author = {Greten, L. and Meyer, S. and Schr{\"o}der, C. and Knorr, A. and Hughes, S.}, title = {Probing the nonlocality of {L}andau levels in {GaAs} quantum wells through modified {P}urcell factors, {L}amb shifts and dipole emitted spectra}, year = {2026}, eprint = {2607.21528}, archivePrefix = {arXiv}}

@article{RomanRoche2021, author = {Rom{\'a}n-Roche, J. and Luis, F. and Zueco, D.}, title = {Photon condensation and enhanced magnetism in cavity {QED}}, journal = {Phys. Rev. Lett.}, volume = {127}, pages = {167201}, year = {2021}, doi = {10.1103/PhysRevLett.127.167201}}

@article{Dicke1954, author = {Dicke, R. H.}, title = {Coherence in spontaneous radiation processes}, journal = {Phys. Rev.}, volume = {93}, pages = {99}, year = {1954}}

@article{HeppLieb1973, author = {Hepp, K. and Lieb, E. H.}, title = {On the superradiant phase transition for molecules in a quantized radiation field: the {D}icke maser model}, journal = {Ann. Phys. (N.Y.)}, volume = {76}, pages = {360}, year = {1973}}

@article{WangHioe1973, author = {Wang, Y. K. and Hioe, F. T.}, title = {Phase transition in the {D}icke model of superradiance}, journal = {Phys. Rev. A}, volume = {7}, pages = {831}, year = {1973}}

@article{EmaryBrandes2003, author = {Emary, C. and Brandes, T.}, title = {Quantum chaos triggered by precursors of a quantum phase transition: the {D}icke model}, journal = {Phys. Rev. Lett.}, volume = {90}, pages = {044101}, year = {2003}}

@article{Baumann2010, author = {Baumann, K. and Guerlin, C. and Brennecke, F. and Esslinger, T.}, title = {Dicke quantum phase transition with a superfluid gas in an optical cavity}, journal = {Nature}, volume = {464}, pages = {1301}, year = {2010}}

@article{GarciaVidal2021, author = {Garcia-Vidal, F. J. and Ciuti, C. and Ebbesen, T. W.}, title = {Manipulating matter by strong coupling to vacuum fields}, journal = {Science}, volume = {373}, pages = {eabd0336}, year = {2021}}

@article{Schlawin2022, author = {Schlawin, F. and Kennes, D. M. and Sentef, M. A.}, title = {Cavity quantum materials}, journal = {Appl. Phys. Rev.}, volume = {9}, pages = {011312}, year = {2022}}

@article{Bloch2022, author = {Bloch, J. and Cavalleri, A. and Galitski, V. and Hafezi, M. and Rubio, A.}, title = {Strongly correlated electron--photon systems}, journal = {Nature}, volume = {606}, pages = {41}, year = {2022}}

@article{ParaviciniBagliani2019, author = {Paravicini-Bagliani, G. L. and others}, title = {Magneto-transport controlled by {L}andau polariton states}, journal = {Nat. Phys.}, volume = {15}, pages = {186}, year = {2019}}

@article{MazzaGeorges2019, author = {Mazza, G. and Georges, A.}, title = {Superradiant quantum materials}, journal = {Phys. Rev. Lett.}, volume = {122}, pages = {017401}, year = {2019}}

@article{Guerci2020, author = {Guerci, D. and Simon, P. and Mora, C.}, title = {Superradiant phase transition in electronic systems and emergent topological phases}, journal = {Phys. Rev. Lett.}, volume = {125}, pages = {257604}, year = {2020}}

@article{Bastidas2012, author = {Bastidas, V. M. and Emary, C. and Regler, B. and Brandes, T.}, title = {Nonequilibrium quantum phase transitions in the {D}icke model}, journal = {Phys. Rev. Lett.}, volume = {108}, pages = {043003}, year = {2012}}

@article{OkaKitamura2019, author = {Oka, T. and Kitamura, S.}, title = {Floquet engineering of quantum materials}, journal = {Annu. Rev. Condens. Matter Phys.}, volume = {10}, pages = {387}, year = {2019}}

@article{Bukov2015, author = {Bukov, M. and D'Alessio, L. and Polkovnikov, A.}, title = {Universal high-frequency behavior of periodically driven systems: from dynamical stabilization to {F}loquet engineering}, journal = {Adv. Phys.}, volume = {64}, pages = {139}, year = {2015}}

@article{Hubener2021, author = {H{\"u}bener, H. and De Giovannini, U. and Sch{\"a}fer, C. and Andberger, J. and Ruggenthaler, M. and Faist, J. and Rubio, A.}, title = {Engineering quantum materials with chiral optical cavities}, journal = {Nat. Mater.}, volume = {20}, pages = {438}, year = {2021}}

@misc{Jochl2026, author = {J{\"o}chl, E. and Helmrich, F. and Lindel, F. and Hale, L. and Graziotto, L. and Jarrahi, M. and Nova, T. F. and Faist, J. and Scalari, G.}, title = {Absence of a superradiant phase transition in {D}irac {L}andau polaritons}, year = {2026}, eprint = {2605.27319}, archivePrefix = {arXiv}}

@article{Lin2026, author = {Lin, S.-Y. and Gao, F. and Xu, Y.-J. and Shen, L. and Wang, Y. and Luo, X.-Q. and Zhang, G.-Q.}, title = {Superradiant phase transition in cavity magnonics via {F}loquet engineering}, journal = {Phys. Rev. A}, volume = {114}, pages = {033726}, year = {2026}, doi = {10.1103/kqzn-32hy}}

@misc{HuangTian2022, author = {Huang, J.-F. and Tian, L.}, title = {Superradiant phase transition in the strong coupling regime}, year = {2022}, eprint = {2208.12524}, archivePrefix = {arXiv}}

@article{Johansen2022, author = {Johansen, C. H. and Lang, J. and Morales, A. and Baumg{\"a}rtner, A. and Donner, T. and Piazza, F.}, title = {Multimode-polariton superradiance via {F}loquet engineering}, journal = {SciPost Phys.}, volume = {12}, pages = {094}, year = {2022}, doi = {10.21468/SciPostPhys.12.3.094}}
\end{document}